\documentclass[traditabstract]{aa} 
\usepackage{graphicx} 
\usepackage{natbib} 
\usepackage{txfonts} 
\usepackage{lscape} 

\def\mm{$m_{\mathrm{max}}-M_{\mathrm{cl}}$}

\def\msun{\hbox{$M_\odot$}}

\def\kms{\hbox{km$\,$s$^{-1}$}} 
\def\cm3{\hbox{cm$^{-3}$}} 
\def\ext{H$\alpha$/H$\beta$}

\def\micron{$\mu$m}

\begin{document}

\title{The VLT-FLAMES Tarantula Survey IV:\\
Candidates for isolated high-mass star formation\\
in 30 Doradus}

\author{E. Bressert\inst{1,2,3} 
\and N. Bastian\inst{1,4} 
\and C. J. Evans\inst{5} 
\and H. Sana\inst{6} 
\and V. H\'{e}nault-Brunet\inst{7} 
\and S. P. Goodwin\inst{8} 
\and R. J. Parker\inst{9} 
\and M. Gieles\inst{10} 
\and J. M. Bestenlehner\inst{11} 
\and J. S. Vink\inst{11} 
\and W. D. Taylor\inst{7} 
\and P. A. Crowther\inst{8} 
\and S. N. Longmore\inst{2} 
\and G. Gr\"{a}fener\inst{11} 
\and J. Ma\'{i}z Apell\'{a}niz\inst{12} 
\and A. de Koter\inst{6,13} 
\and M. Cantiello\inst{14} 
\and J.M.D. Kruijssen\inst{13,15,16} }

\institute{School of Physics, University of Exeter, Stocker Road, Exeter EX4 4QL, UK \label{1}\\
\email{ebresser@eso.org} 
\and European Southern Observatory, Karl-Schwarzschild-Strasse 2, D87548, Garching bei M\"{u}nchen, Germany \label{2} 
\and Harvard-Smithsonian CfA, 60 Garden Street, Cambridge, MA 02138, USA\label{3} 
\and Excellence Cluster Universe, Boltzmannstr. 2, 85748 Garching, Germany\label{4} 
\and UK Astronomy Technology Centre, Royal Observatory Edinburgh, Blackford Hill, Edinburgh, EH9 3HJ, UK\label{5} 
\and Astronomical Institute Anton Pannekoek, University of Amsterdam, Kruislaan 403, 1098 SJ, Amsterdam, The Netherlands\label{6} 
\and Scottish Universities Physics Alliance (SUPA), Institute for Astronomy, University of Edinburgh\label{7}\\
Royal Observatory Edinburgh, Blackford Hill, Edinburgh, EH9 3HJ, UK 
\and Department of Physics and Astronomy, University of Sheffield, Sheffield, S3 7RH, UK\label{8} 
\and Institute for Astronomy, ETH Zurich, Wolfgang-Pauli-Strasse 27, 8093, Zurich, Switzerland\label{9} 
\and Institute of Astronomy, University of Cambridge, Madingley Road, Cambridge, CB3 0HA, UK\label{10} 
\and Armagh Observatory, College Hill, Armagh, BT61 9DG, Northern Ireland, UK\label{11} 
\and Instituto de Astrof\'{i}sica de Andaluc\'{i}a-CSIC, Glorieta de la Astronom\'{i}a s/n, E-18008 Granada, Spain\label{12} 
\and Astronomical Institute, Utrecht University, Princetonplein 5, 3584 CC, Utrecht, The Netherlands \label{13} 
\and Kavli Institute for Theoretical Physics, Kohn Hall, University of California, Santa Barbara, CA 93106, USA \label{14} 
\and Leiden Observatory, Leiden University, PO Box 9513, 2300 RA Leiden, The Netherlands \label{15} 
\and Max-Planck Intitut f\"{u}r Astrophysik, Karl-Schwarzschild-Strasse 1, D87548, Garching bei M\"{u}nchen, Germany \label{16} }

\date{Received May 12, 2011; accepted April 12, 2012}

\abstract {Whether massive stars ($\gtrsim 30~\msun$) can occasionally form in relative isolation (e.g. in clusters with $M < 100~\msun$) or if they require a large cluster of lower-mass stars around them is a key test in the differentiation of star formation theories as well as how the initial mass function of stars is sampled. Previous attempts to find O-type stars that formed in isolation were hindered by the possibility that such stars are merely runaways from clusters, i.e., their current isolation does not reflect their birth conditions. We introduce a new method to find O-type stars that are not affected by such a degeneracy. Using the VLT-FLAMES Tarantula Survey and additional high resolution imaging we have identified stars that satisfy the following constraints: 1) they are O-type stars that are not detected to be part of a binary system based on radial velocity (RV) time series analysis; 2) they are designated spectral type O7 or earlier ; 3) their velocities are within 1$\sigma$ of the mean of OB-type stars in the 30 Doradus region, i.e. they are not runaways along our line-of-sight; 4) the projected surface density of stars does not increase within 3~pc towards the O-star (no evidence for clusters); 5) their sight lines are associated with gaseous and/or dusty filaments in the interstellar medium (ISM), and 6) if a second candidate is found in the direction of the same filament with which the target is associated, both are required to have similar velocities. With these criteria, we have identified 15 stars in the 30 Doradus region, which are strong candidates for being high-mass stars that have formed in isolation. Additionally, we employed extensive Monte Carlo stellar cluster simulations to confirm that our results rule out the presence of clusters around the candidates. Eleven of these are classified as Vz stars, possibly associated with the zero-age main sequence. We include a newly discovered Wolf-Rayet star as a candidate, although it does not meet all of the above criteria.}

\maketitle

%
\section{Introduction} \label{sec:intro}

Massive stars play a crucial role in shaping their environment by ionising large regions around them, affecting the temperature and structure of the ISM and through chemical enrichment. They are the flag-posts of star formation, with large H~{\sc ii} regions often dominating the optical structure of their host galaxies, e.g., 30 Doradus in the Large Magellanic Cloud (LMC). Despite their prominence, we know little about how massive stars form due to multiple challenges: their large distances to us, heavy extinction, rapid formation process, and small population statistics compared to their lower mass counterparts.

The spatial distribution of massive stars can provide us with clues to their formation. Do they only form in massive clusters or can they also form in isolation? There are multiple scenarios on how massive stars form and here we consider two of them, as these represent opposite ends of the spectrum. The first is competitive accretion (e.g., \citealt{Bonnell2001,Bonnell2004,smith2009}) where stars, including massive ones, only form in clustered environments. In this model, massive stars are built up from the seeds of lower mass stars through the accretion of gas from their environment. Since a large amount of gas is needed for accretion, a cluster of low mass stars is expected to be present around the higher mass stars (e.g., ~\citealt{Maschberger2010}). The second model of massive star formation is that of monolithic collapse (\citealt{Yorke2002,Mckee2003,Krumholz2009}) where a star's mass is set by the initial dense core from which it will form. In this scenario, massive stars can form, albeit rarely, without a surrounding cluster. Hence, determining the spatial distributions of the massive stars allows us to test these theories and will possibly lead to a better understanding of the star formation process.

In addition to testing star formation theories, determining whether or not massive stars can form in isolation would have important ramifications on how the stellar initial mass function (IMF) is sampled (e.g. \citealt{Bastian2010}). Depending on how stars form they will sample the IMF in different ways. It has been suggested that in the competitive accretion scenario massive stars only form in clusters, so the IMF will be sampled in a ``sorted'' way, such that enough low mass stars need to be present before higher mass stars can form (\citealt{Weidner2010}). However, in the monolithic collapse scenario, massive stars are able to form in relative isolation, meaning that the IMF will be sampled stochastically (e.g., \citealt{Oey2004,Elmegreen2006a,Parker2007,Selman2008}), likely reflecting the mass distribution of the star-forming cores (e.g., \citealt{Alves2007}). These different ways of sampling the IMF have important implications to the resulting mass distributions, and in particular massive star numbers, in different galaxies (e.g. \citealt{Weidner2010,Bastian2010}; also see Sect. \ref{sec:summary}).

These two IMF sampling scenarios lead to similar relationships between the mass of the most massive star in a group and the mass of the group in total. However, there are important differences. In the stochastic scenario, such a relation is statistically expected (with a large scatter), whereas in the sorted sampling scenario such a relationship is causally expected with little scatter. Observations have been presented which seem to favour one scenario or another (e.g., \citealt{Maschberger2008,Weidner2010}), however large homogeneous datasets are required to provide a solid answer. Alternatively, in the present work, we will search for extremes in the distribution, O-type stars that formed in relative isolation, in order to address this issue.

The term `isolated massive star formation' does leave room for different interpretations and one could argue against a star being `isolated' in several different ways, which we list below, from most to least restrictive.
\begin{enumerate}
	\item The massive stars formed from the same molecular cloud or filament (i.e. the stars would be associated with the same OB associations). 
	\item The massive stars formed in the same or a related environment in terms of radiation field or kinetic energy input (e.g., a second massive star formed due to feedback from a first star) 
	\item The massive stars formed within the same gravitational well of size $\lesssim$3~pc that led them to be (at least initially) bound and constitute a physical cluster \citep[see ][]{Efremov1998,Maiz-Apellaniz2001,Scheepmaker2007} and not simply part of an unbound association. 
\end{enumerate}

\noindent In general terms, case 1 is responsible for large-scale similarities in age, location, and composition but is otherwise irrelevant for the detailed physics of star formation. Case 2 influences the overall efficiency and the age distribution within a large region and possibly the IMF, since triggering can, in principle, lead to variations. In case 3, only within the dense (sub)parsec scale regions, can gravitational interactions between different clumps and/or stars happen within the massive-star formation timescale of $\sim10^5$ years \citep{Mckee2003}. In other words, from the gravitational point of view, it is mostly irrelevant whether a large mass (molecular cloud or other clusters) is located at a distance of 10 pc because for a period of $10^5$ years this would constitute a nearly constant gravitational field. It is only nearby ($\sim$1 pc) parts of the cloud that have orbital time scales short enough to lead to multiple interactions, such as accretion, close encounters, and collisions. {\it Therefore, from the point of view of star formation theories and the origin of the IMF, case 3 is the relevant scale, which is what we will adopt in the current paper.} 

The possible effect of triggering (case 2) could work as an external agent to influencing massive star formation in 30 Doradus. The central cluster, R136, other clusters and massive stars have been suggested to influence the massive star formation process in the region \citep[see ][]{Brandner2001,Walborn1999,Walborn2002}, but \cite{DeMarchi2011} questions whether triggered star formation is relevant as their investigation with the {\it Hubble Space Telescope's} Wide Field Camera 3 (WFC3) observations do not show clear evidence of such causal effects. In either scenario, whether triggering (case 2) is influencing massive star formation or not in 30 Doradus, it would not change a massive star's candidacy for forming in isolation following the case 3 definition. Once triggering activates star formation it is up to the parts of the cloud within the cluster length scale to interact with each other.

When we talk about massive stars and them forming in ``isolation'', we adopt a similar \footnote{Parker \& Goodwin (2007) define an isolated massive star ($\geq$17.5 \msun~-- a criterion that is not used in this paper) as one with a surrounding stellar cluster with a total mass of $<100$~\msun~and which does not contain any early B-type stars.} definition given in \cite{Parker2007}, where massive stars refer to those those with spectral types of O7 or earlier ($\gtrsim 30$ \msun~depending on mass estimate method used -- discussed later on in the paper) and isolation means that they are not found in clusters of $\geq 100~\msun$ and $r\lesssim3$~pc. Once massive stars are located, we will estimate the expected underlying cluster mass according to the \citep{Weidner2010} theory by using the \mm\ relation, which is expected to hold in the sorted sampling scenario. This allows us to assess if the underlying cluster should have been detected, and if the O-type stars are truly isolated, will enable constraints to be put on star formation theories as well as IMF sampling scenarios.

We are not the first to look for massive stars forming in isolation. A comprehensive analysis of isolated field O-type stars (hereafter referred to as O-stars) in our Galaxy was conducted by \cite{deWit2004,deWit2005} who found that 4 $\pm$ 2 percent of the O-stars in their sample (model derived value from observations) could not be traced back to clusters and hence, likely formed in isolation. Similarly, \cite{Lamb2010} looked at the Small Magellanic Cloud (SMC) and found three O-stars that are in sparse-clusters. The total mass of the sparse-clusters in relation to the mass of the O-stars is not compatible with the sorted sampling scenario. The sparse-clusters fall within the definition as we defined above for ``isolated''. The SMC was also investigated for isolated star formation by \cite{Selier2011}, where they reported an interesting compact H$\,{\rm II}$ region, N33. The compact H$\,{\rm II}$ region cannot be traced back to any nearby clusters, associations or molecular clouds and may be evidence for isolated massive star formation at the earliest stages. Here we consider observations from the VLT-FLAMES Tarantula Survey (hereafter VFTS; Evans et al. 2011) of the 30 Doradus nebula and its environs which contains ~350 O-stars. 

In addition to the de Wit et al. (2004,2005) findings, there have been several theoretical investigations of the likelihood of isolated, massive star formation. Firstly, \cite{Oey2004} noted that the number of stars per cluster appears to follow a power-law like slope. They conclude that the power-law dependence extends, continuously, all the way to down to one OB star per group, association or cluster, i.e. $N_{\star} = 1$. Parker \& Goodwin (2007) simulated a large number of stellar clusters assuming a cluster mass function with $\beta = 2$ and a universal IMF (stochastic sampling). They found that ``isolated'' massive star formation is not unexpected in this scenario, and that contrary to \cite{Weidner2006} who found a strict \mm\ relation, stochastic sampling leads to a \mm\ relation with significantly more scatter and a slightly different slope. In the sorted sampling scenario, the scatter is significantly reduced, due to the explicit link between the cluster mass and the stars that it forms. However, without a large number of clusters, it is difficult to differentiate between the two \citep{Maschberger2008}. The two scenarios do differ significantly in the extreme end of the distributions, i.e. the presence/absence of isolated high mass stars. Lamb et al. (2010) conducted similar and updated simulations as presented by Parker \& Goodwin (2007) and confirmed their results.

It is important to note that Weidner \& Kroupa~(2006) and Weidner et al.~(2010) adopt similar scales from case 3 (concerning arguments against `isolated' stars) for defining the \mm\ relation, i.e., they adopted the cluster length scale and not the OB association scale. For example, the authors use the Orion Nebular Cluster (ONC) and not the entire Orion complex (an OB association) for supporting the \mm\ relation. Using the cluster scale makes sense if all stars are formed in clusters, i.e., in quantised units of star formation. However, if star formation is hierarchical (e.g., Elmegreen et al~2006) then no distinct scale exists in the star formation process (above the scale of individual stars/cores). This means that clusters are made up of sub-clusters which merge as they evolve dynamically, and that clusters/associations are themselves grouped into larger structures (e.g., \citealt{bastian2005}). In a hierarchical scenario, the individual sub-clusters cannot fit the \mm\ relation {\it if} the final cluster does (since the most massive star would be too massive for its sub-cluster size), so the stars must `know' about the cluster that they will finally be a part of. This same argument also holds for larger associations and cluster complexes. Hence, the \mm\ relation must have a length scale associated to it, if star formation is hierarchical, above and below which the relation breaks down.

The 30 Doradus region is a complex and dynamic region that contains multiple, although not necessarily spatially distinct, generations of stars (see \citealt{Walborn1997}). The youngest population is dominated by the central cluster R136, with an age of 1-2 Myr \citep[e.g. ][]{deKoter1998,Massey1998} and, most pertinently for the discussion here, there appears to be another young population to the north and west of R136, exemplified by the compact multiple systems in the dense nebular knots observed with the Hubble Space Telescope by \cite{Walborn1999,Walborn2002}. These comprise an apparently young, still embedded phase of star formation. Interestingly, \cite{Walborn2002} also presented imaging of two notable infrared (IR) sources, one of which was resolved into a small, embedded cluster, while the other was a point like source, seemingly single monolithic object (albeit at the distance of the LMC). This led the authors to note that the later object may have formed without an associated cluster or association.

R136 has a relatively shallow power-law density profile and it does not appear to have a strong truncation out to at least 10 pc (e.g. \citealt{Campbell2010}). All massive stars observed beyond this radius are likely to be either runaway stars, or to have formed {\it in situ}. The majority of the stars in our sample are young (≤ 2 Myr), hence the disruption of large clusters with lower-mass stars blending into the background due to ``infant mortality'', the ``cruel cradle effect''\footnote{While infant mortality indicates the early disruption of stellar structure by gas expulsion, the cruel cradle effect refers to the disruptive influence of tidal perturbations by the dense, star-forming environment, which act on a similar timescale.} or the dispersal of unbound associations (either of which would cause the most massive star to appear isolated), is not expected to influence our analysis (e.g. \citealt{Bastian2006,Gieles2011,Kruijssen2011,Kruijssen2012,Girichidis2012}).

In the present work, we attempt to overcome the limitations of the previous studies by removing the possibility that the stars are runaways by using multi-epoch medium-resolution spectroscopy and by cross-correlating the spatial distribution of the candidates with known gaseous filaments. We do this by combining observations of 30 Doradus with several instruments from the {\it Very Large Telescope} (VLT), {\it Hubble Space Telescope} (HST), and the {\it Spitzer Space Telescope} (SST), which will be discussed in Sect. \ref{sec:obs}. The methods to finding isolated massive star candidates are discussed in Sect. \ref{sec:method}, where the results and conclusions are discussed in Sect. \ref{sec:res}, Sect. \ref{sec:conclusions}. In the final section we provide a summary and discuss implications.

\section{Observations} \label{sec:obs}
\begin{figure}
	\resizebox{\hsize}{!}{
	\includegraphics{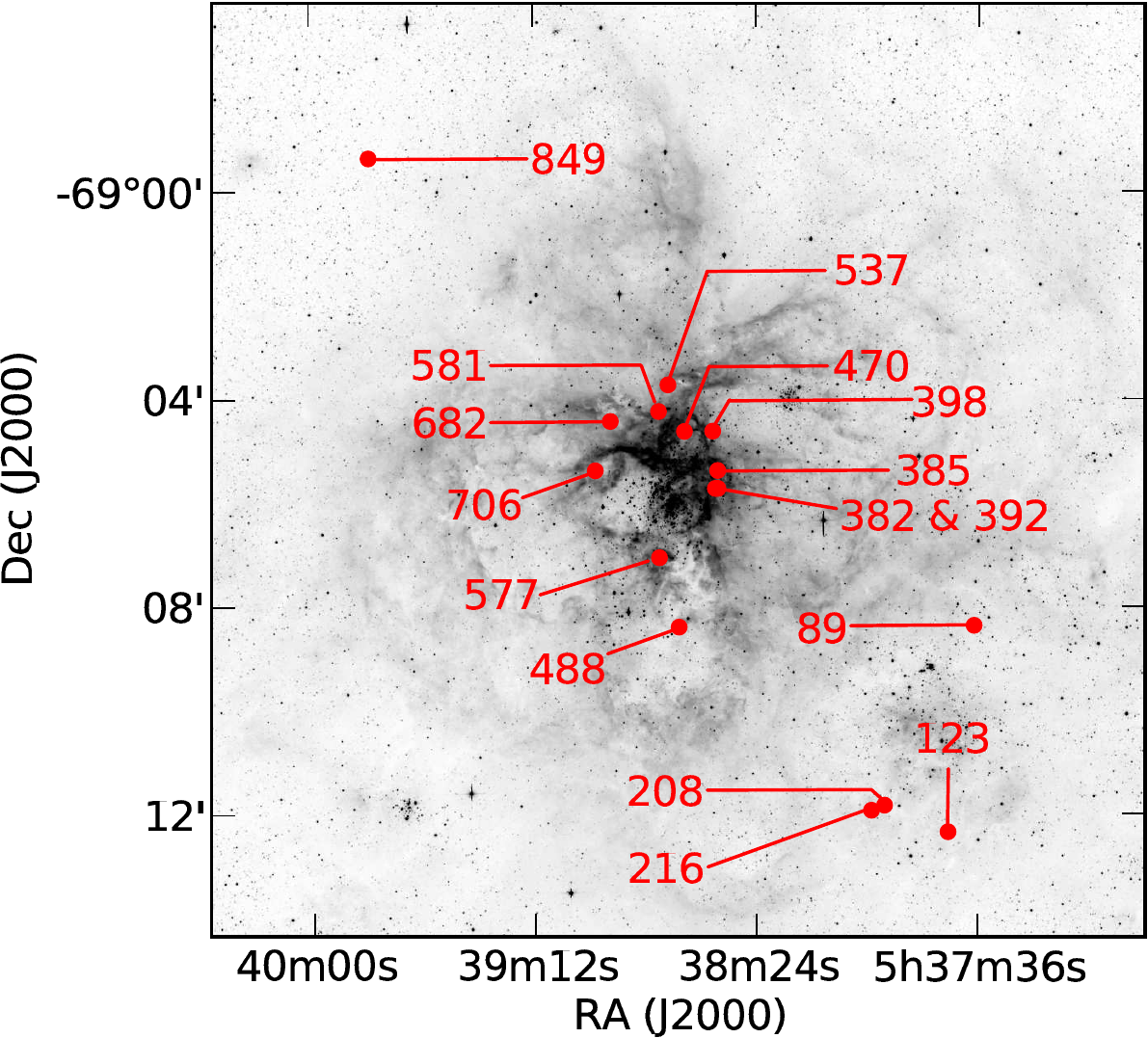}} \caption{The 16 candidates, isolated massive stars, in 30 Doradus are marked with red circles. Figs. \ref{fig:fig05}--\ref{fig:fig09} show sub-fields of this region to highlight gas/dust filamentary structures that are likely to be associated with the massive stars in question. The stars are identified with the numbers used for the VLT FLAMES Tarantula Survey.} \label{fig:fig01} 
\end{figure}

In this work we used optical spectroscopy and optical to mid-IR imaging to study the O-stars in 30 Doradus. The core dataset is composed of medium resolution optical spectra from the VLT-FLAMES Tarantula Survey (Evans et al. 2011). bf VFTS allows us to identify the spectral type of each star, test whether the star is part of a spectroscopic binary and measure each star's radial velocity. The survey employs three different modes of the Fibre Large Array Multi-Element Spectrograph (FLAMES; \citealt{Pasquini2002}) instrument on the VLT: {\it Medusa-Giraffe}, {\it ARGUS-Giraffe}, and the {\it Ultraviolet and Visual Echelle Spectrograph}. In this paper we only use spectra that were observed using the Medusa fibre-feed to the Giraffe spectrograph. There are 132 fibres available for observations, deployable within a $25'$ diameter field-of-view and with a diameter of $1\farcs 2$ on the sky. The {\it European Southern Observatory} (ESO) Common Pipeline Library (CPL) FLAMES reduction routines (v2.8.7) were used for all of the data processing. The standard reductions were then applied such as Heliocentric correction and sky subtraction. Additional information regarding the observations, reductions and survey strategy are given in Evans et al. (2011). The radial velocity and multiplicity analysis is discussed in Sana et al. (in prep.).

The second part of the dataset is composed of imaging surveys to 1) find filamentary structures of gas and dust, i.e., the sites of star formation and 2) determine if a given star is truly isolated. For the first goal we use the \ext~map from \cite{Lazendic2003}, 70 \micron~maps from the SST Legacy survey Surveying the Agents of Galaxy Evolution (SAGE) \citep{Meixner2006}, and VLT High Acuity Wide field K-band Imager (HAWK-I) observations \citep{Kissler-Patig2008} where ionised filaments can be detected due to the presence of the Br$\gamma$ emission line within the bandpass. To determine whether there are significant clusters associated with the massive stars we use HST imaging with the Wide-Field Planetary Camera 2 (WFPC2) and the Advanced Camera for Surveys (ACS) instruments as well as the VLT HAWK-I K$_s$ band images to search for embedded clusters. 

In Fig.~\ref{fig:fig01} we show a V-band image of 30 Doradus taken with ESO and the {\it Max Planck Gesellschaft} (MPG) 2.2m telescope using the Wide Field Imager (WFI) instrument \citep{Baade1999} to show the 30 Doradus region for context (Program ID: 076.C-0888; PI J. Alves). 

\begin{figure}
	\resizebox{\hsize}{!}{
	\includegraphics{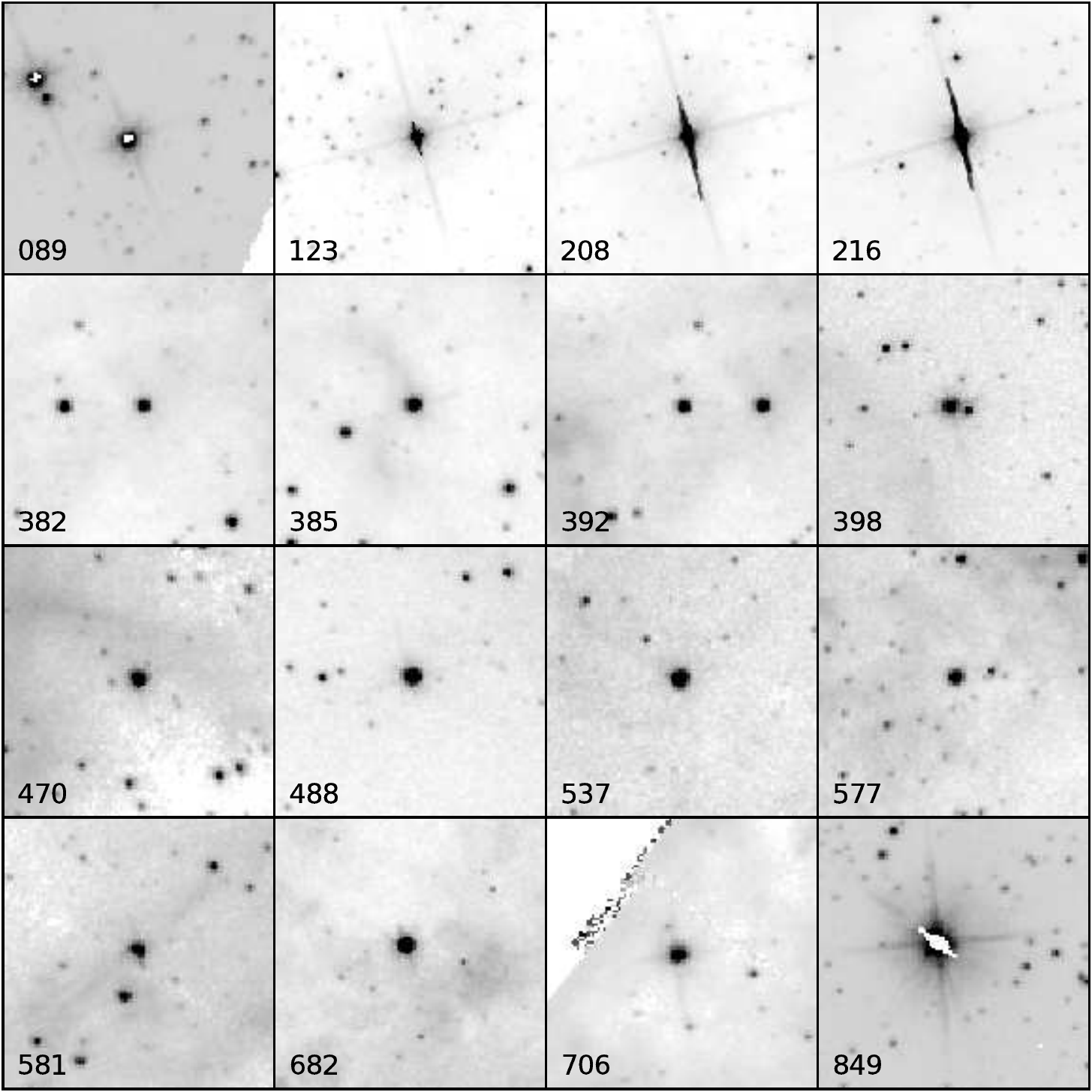}} \caption{2.5$\times$2.5 parsec (10''$\times$10'') logarithmically stretched grey scale images of each of the stars. All of the candidates were observed in the F814W band except for VFTS 682 and VFTS 849, which are observed in the F673N and F606W bands, respectively. Additionally, all of the candidates were observed with the WFPC2, except for VFTS 123, 208 and 216 which were observed with ACS.} \label{fig:fig02} 
\end{figure}

For the candidates, except VFTS 089 and 849, we acquired the HST data from the Hubble Legacy Archive, and selected the science grade data for analysis. The proposal ID numbers for the HST data are: 05114, 08163, 09471. For VFTS 089 and 849 we used high-level science products from The Archival Pure Parallels Project (APPP) on the LMC. \cite{Wadadekar2006} describes the data processing and quality of the HST images. WFPC2 and ACS data typically have pixel resolutions of 0.1'' and 0.05'', respectively. See Fig. \ref{fig:fig02}~for 2.5$\times$2.5 pc subplots of the HST data for each candidate. Further details on HST's sensitivity around the O-star candidates is discussed in Sect. \ref{sec:scm}.

\section{Method} \label{sec:method} To determine that an O-star is a not a runaway and is still located near its birth-site, we need to apply several constraints on the star in the line-of-sight and the plane of the sky. Several observational tools are needed to do this: precise radial velocity measurements (line-of-sight), association with filamentary structure (plane of sky), and relative isolation from lower mass stars.

The radial velocity measurements of all the single O-stars in VFTS (Sana et al. in prep.) can be well fit by a gaussian distribution. The mean radial velocity is 270.73 $\kms$~with a dispersion (1$\sigma$) of 10.50 $\kms$. These values are used in selecting the candidates (see criterion 3).

\subsection{Criteria} \label{sec:critera} We begin with the VFTS observations of the 30 Doradus region. The survey will provide multi-epoch radial velocity measurements for a sample of $\sim800$ OB-type stars in the 30 Doradus region with a typical precision better than 5 $\kms$. From the sample we select candidates based on the selection criteria provided below. Within brackets we report on the number of stars that fulfil all successive criteria in the order as listed and which constraint it supports (e.g. plane-of-sky, line-of-sight, or isolation). The criteria are:
\begin{enumerate}
	
	\item O-type stars that do not show significant variations in their RV, meaning that they are unlikely part of a spectroscopic binary system. This allows us to determine accurately the systemic velocity of each star. In Sect. \ref{sec:binarity}, we quantitatively estimate our detection biases and we will show that our sample cannot be heavily contaminated by undetected spectroscopic binaries. (stars: 184, which include some B0-type stars from preliminary classifications) (constraint: line-of-sight)
	
	\item Their sub-spectral type are O7 or earlier, which corresponds so stellar masses of $\gtrsim 30$ \msun. See Sect. \ref{sec:spectral} for details on mass estimates. (stars: 65).
	
	\item Their radial velocities are within one $\sigma$ of the mean of all the O-stars in the VFTS sample of 30 Doradus. This ensures that the stars are not runaways in the line-of-sight. (stars: 39) (constraint: line-of-sight).
	
	\item That the candidate star be located on (projected) gaseous filament seen in either ionised gas (i.e., H$\alpha$ or Br$\gamma$), warm dust (i.e., 70 \micron), or cold gas/dust (i.e., based on extinction or molecular maps). These gaseous filaments are the likely sites of star formation, and the chance projection of all ejected stars from nearby clusters that lie projected upon these filaments is exceedingly low. (stars: 27) (constraint: plane-of-sky)
	
	\newcounter{enumi_saved} \setcounter{enumi_saved}{\value{enumi}} 
\end{enumerate}

In addition to the above criteria, we use high resolution HST imaging (when available) of the regions around the candidate stars, along with deep ground-based near-infrared imaging from VLT to place constraints on the size of any potential cluster surrounding the candidate O-star. With this data we add the following criterium:
\begin{enumerate}
	\setcounter{enumi}{\value{enumi_saved}} 
	\item The surface density distribution of stars does not increase {\it towards the candidate star} within 3~pc, see Fig.\ ~\ref{fig:fig04}. Note that archival HST data does not cover all candidates from criterion 4. Due to the lack of HST data, four sources are automatically rejected from consideration. This means that this step automatically reduced the number of stars from the previous step of 27 stars to 23 stars. Then we apply the surface density criterion on the remaining 23 stars. (stars: 15) (constraint: isolation)
	
	\newcounter{enumi_saved3} \setcounter{enumi_saved3}{\value{enumi}} 
\end{enumerate}

\noindent Where applicable, if two stars are located along the same gaseous filament, we apply the additional criterion below. Note, that the additional criterion does not exclude the 15 candidates from above.
\begin{enumerate}
	\setcounter{enumi}{\value{enumi_saved3}} 
	\item Two stars have radial velocities within 5 $\kms$~of each other. While filaments can have flows along them, we expect the gas/stars within a filament to have very similar radial velocities. (stars: 2) (constraints: plane-of-sky, line-of-sight) 
\end{enumerate}

With these criteria we can build up a collection of candidates that may have formed in isolation. Once a homogeneous map of 30 Doradus showing the gas/dust filamentary structures become available, e.g. \emph{Herschel Space Observatory} maps (\citealt{Meixner2010}) or extinction maps derived from the Visible and Infrared Survey Telescope for Astronomy (VISTA) Magellenic Cloud Survey \citep{Cioni2011}, we will be able to apply simple probability tests to strongly limit the possibility of massive stars being runaways (see Appendix \ref{sec:appendix}).

\subsection{Spectral types and ages} \label{sec:spectral}

The ages of our isolated O-star candidates likely range from less than 1 Myr to more than 4 Myr based on their spectral types (Weidner \& Vink 2010). We have therefore assigned \emph{grades} to each star on the basis of their spectral types, employing estimated ages from Weidner \& Vink (2010) for LMC stars. Grade 1 candidates are most likely 2 Myr old or younger, grade 2 are between 2 and 4 Myr, and Grade 3 are those older than 4 Myr. The grade scheme is provided to highlight possible issues of candidate associations, depending on their age, with filamentary gas/dust structure (see Sect. \ref{sec:filaments}).

\section{Results} \label{sec:res}

The properties of the 15 isolated O-star candidates are given in Table~\ref{tab:tab01}, including their positions, mean radial velocities (of the gas and stars), spectral types (Walborn et al. in prep), masses (estimated from the spectral types), V-band magnitudes, their expected cluster masses based on the \mm\ relation, and a grade to indicate their likely ages. Eleven of the candidates are classified as Vz stars, deﬁned by stronger He\,{\sc ii}~$\lambda$4686 absorption than either He\,{\sc i}~$\lambda$4471 or He\,{\sc ii}~$\lambda$4542 in their spectra \citep{Walborn1992,Walborn1997}. The Vz phenomenon has been proposed to be related to stars located close to the zero-age main sequence (ZAMS). Alternatively, rotational broadening may induce an apparent modification of the line strength and of He line ratios. In the latter case the Vz signature may be common for normal dwarf stars and not be tracing stars close to the ZAMS (Sab\'{\i}n-Sanjuli\'an, priv. comm.). Quantitative spectroscopic analyses are needed to elucidate the true nature of the Vz signature and will be undertaken for the Vz stars in the VFTS sample.

Eleven of the candidates were classified as Vz stars, defined by strong He~{\sc ii} $\lambda$4686 absorption in their spectra \citep{Walborn1992,Walborn1997} and thought to be close to the zero-age main sequence (ZAMS). The ZAMS classification is not absolute since the morphological hypothesis has not yet been confirmed by systematic quantitative analyses; determinations of their gravities and luminosities is underway as part of the VFTS.

The candidates are located at projected distances of between 14 and 130 pc from R136 (see Fig.~\ref{fig:fig01}). Twelve are associated with the active star forming environment around R136 (see Fig.~\ref{fig:fig05} for example). Four are located to the southwest of R136 and are closer to the older association NGC\,2060, which has an associated supernova remnant \citep{Danziger1981,Chu1992}. Note that the distance between the centre of NGC 2060 and the closest candidate in this paper is $>20$~pc.

We have taken the spectral types of the candidates and estimated their respective masses from \cite{Weidner2010c}, for rotating O-star models in the LMC to approximate the expected O-star parent cluster's mass based on \mm. The stellar masses are also used as an input in the computation of the sensitivity of our data to spectroscopic binaries (see Sect. \ref{sec:binarity}). Estimating stellar masses from spectral types is known to suffer from a number of caveats \citep[e.g. ][]{Weidner2010c}, however the masses are precise enough for our objectives (typical errors are expected to be less than 30\%). In the following subsections we discuss the candidates in more detail.
\begin{figure}
	\resizebox{\hsize}{!}{
	\includegraphics{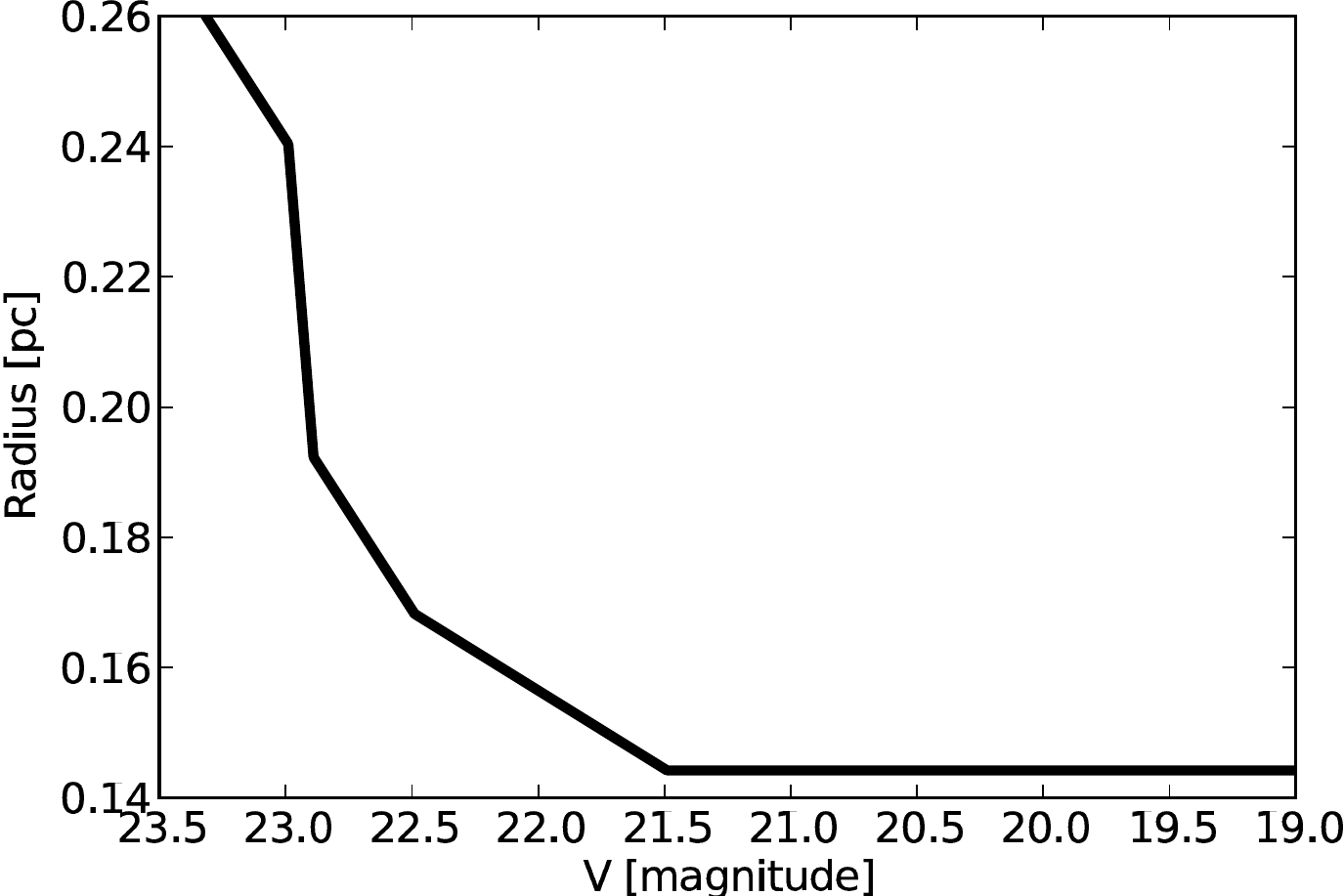}} \caption{The 90\% completeness magnitude limit as a function of radius for VFTS 385 (using HST/WFPC2). The closest detectable source to the candidate is $\sim$ 0.15 pc, where the upturn begins. VFTS 385 is the brightest observed candidate presented in this paper.} \label{fig:fig03}

\end{figure}
\begin{table*}
	\caption{The 16 candidates that most likely formed in isolation.} \centering 
	\begin{tabular}
		{lcccccclccc} Star& Aliases&$\alpha$(2000)&$\delta$(2000)&RV$_{\textrm{star}}$&RV$_{\textrm{ISM}}$&$m_V$&Spectral Type& $M_{\textrm{star}}$& $M_{\textrm{cl}}$& Grade\\
		VFTS& & & & [$\kms$]& [$\kms$]& & & [\msun] & [\msun] & \\
		\hline 089&ST 1-25&05 37 36.87&$-$69 08 22.82&$280.1 \pm 0.8$&$270.6 \pm 2.4$&16.08&O6.5 V((f))z&33&660&1\\
		123$^\dagger$&-&05 37 42.45&$-$69 12 21.58&$270.7 \pm 0.9$&$270.6 \pm 1.1$&15.78&O6.5 Vz&33&660&1\\
		208&ST 1-93&05 37 56.23&$-$69 11 50.90&$270.2 \pm 1.0$&$275.6 \pm 1.7$&14.65&O6 (n)fp&46&1060&3\\
		216&ST 1-97&05 37 59.06&$-$69 11 56.83&$269.4 \pm 0.5$&$274.0 \pm 0.6$&14.41&O4 V((fc))&53&1350&2\\
		382$^\dagger$&S 226&05 38 32.28&$-$69 05 44.57&$278.3 \pm 1.0$&$264.4 \pm 8.8$&15.88&O4-5 V((fc))z&48&1130&1\\
		385&P 288, S 84&05 38 32.32&$-$69 05 23.87&$270.8 \pm 0.6$&$281.3 \pm 0.9$&14.65&O4-5 V((n))((fc))&48&1130&2\\
		392&S 268&05 38 32.83&$-$69 05 44.60&$278.3 \pm 1.0$&$282.1 \pm 1.4$&16.10&O6-7 V((f))z&33&660&1\\
		398&Mk~59, P 341&05 38 33.38&$-$69 04 38.39&$268.8 \pm 0.5$&$274.1 \pm 0.8$&14.40&O5.5 V((n))((f))z&40&860&1\\
		470&P 716&05 38 39.49&$-$69 04 38.64&$265.0 \pm 1.2$&$278.3 \pm 3.4$&15.46&O6-7 V((f))z&33&660&1\\
		488&P 791&05 38 40.72&$-$69 08 24.90&$270.1 \pm 1.0$&$268.0 \pm 2.2$&15.87&O6 V((f))z&36&740&1\\
		537&P 1022&05 38 43.02&$-$69 03 44.78&$271.0 \pm 1.1$&$259.7 \pm 1.5$&15.99&O5 V((fc))z&44&990&1\\
		577$^\dagger$&P 1189&05 38 44.94&$-$69 07 04.59&$264.5 \pm 1.3$&$271.4 \pm 2.8$&16.64&O6 V((fc))z&36&740&1\\
		581&P 1218&05 38 45.07&$-$69 04 15.57&$277.6 \pm 1.0$&$259.1 \pm 3.8$&16.07&O4-5 V((fc))&48&1130&2\\
		682$^{\star ~ \dagger}$&P 1732&05 38 55.51&$-$69 04 26.72&$300.0 \pm 10.0$&$259.5 \pm 3.0$&16.08&WN5h&$>$100&$>$3900&1\\
		706&P 1838&05 38 58.76&$-$69 05 23.93&$269.8 \pm 3.6$&$290.7 \pm 0.6$&15.77&O6-7 Vnnz &33&660&1\\
		849$^\dagger$&-&05 39 47.36&$-$68 59 21.99&$260.6 \pm 0.8$&$262.5 \pm 1.3$&15.14&O7 Vz&30&580&1\\
		\hline 
	\end{tabular}
	\begin{flushleft}
		{{\bf Notes.} Positions and magnitudes come from Evans et al. (2011). Eleven of the 16 candidates are Vz stars, thought to be zero-age main sequence stars (ZAMS). The final column reflects the potential ages of the candidates on the basis of their spectral types and results from Weidner \& Vink (2010). Grade 1 candidates are most likely 2 Myr old or younger, grade 2 are between 2 and 4 Myr, and grade 3 are those older than 4 Myr. The values for $M_{\textrm{cl}}$ are derived using the \mm\ relationship from \cite{Weidner2010}. Errors provided for the radial velocity are uncertainties from the mean. Candidates VFTS 208 and 385 show slight indications of variability, but it does not affect the current results. References. Aliases/previous identifications of the VFTS candidates are given in the second column. The sources of identification are: Mk \citep{Melnick1985}, P \citep{Parker1993}, S \citep{Selman1999} and ST \citep{Schild1992}.}\\
		$^{\star}${\small The source is a Wolf-Rayet star and does not fit all criteria in Sect. 3.1. For further details see Sect. 4.1.4 and Bestenlehner et al. 2011}\\
		$^{\dagger}${\small No known spectroscopy prior to VFTS}
	\end{flushleft}
	\label{tab:tab01} 
\end{table*}
\begin{figure}
	\resizebox{\hsize}{!}{
	\includegraphics{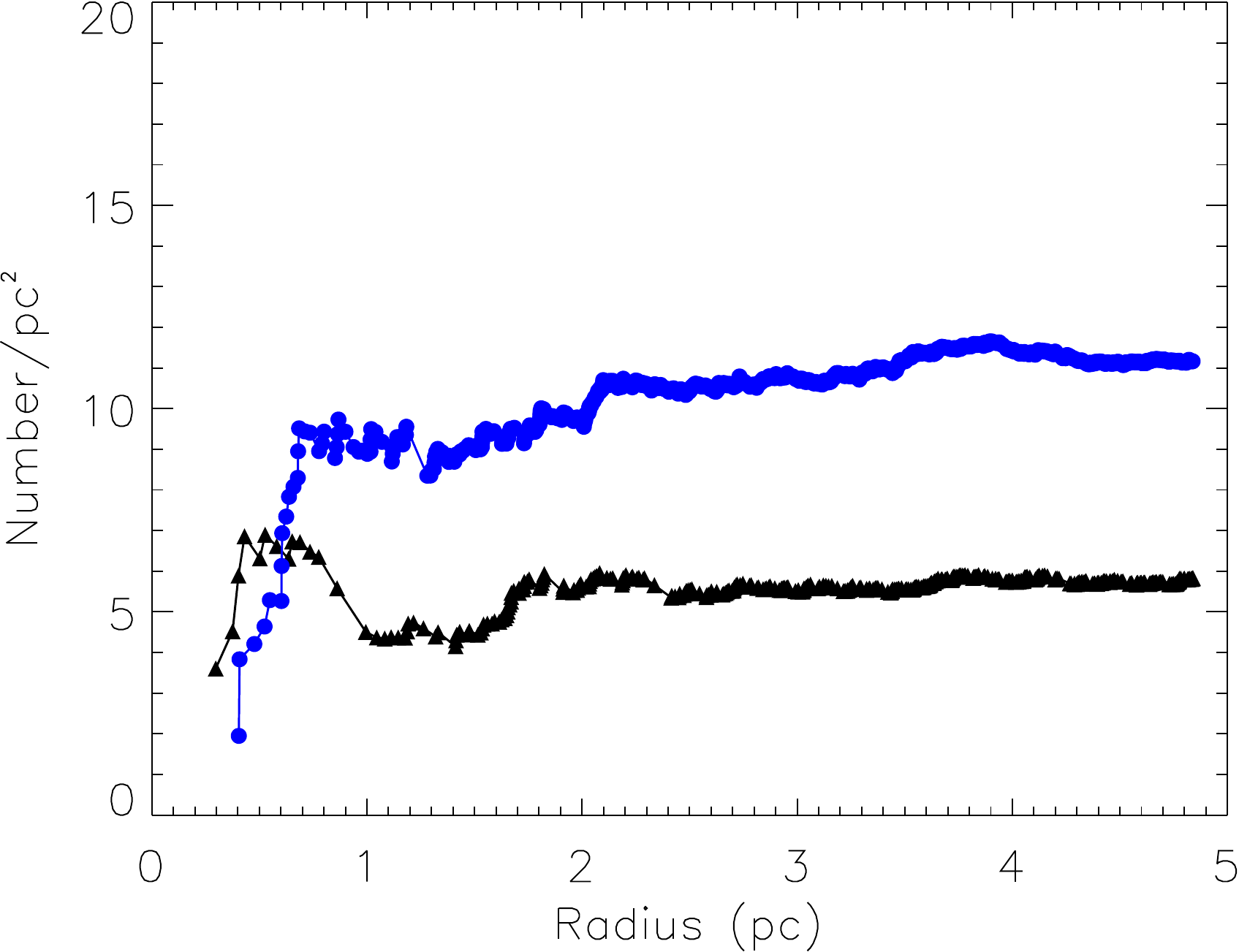}} \caption{HST stellar surface density distributions (cumulative - see \citealt{Lamb2010}) around VFTS 208. The black line with triangles is the HST F555W band and the blue line with circles is the HST F814W band. Due to the saturation on the ACS images caused by the candidate (candidate reference point is at 0 pc), VFTS 208 represents a worst case. However, even here we see that there is no stellar count increase from 3 pc and inwards (to 0~pc) toward the candidate. This has been observed similarly for each candidate.} \label{fig:fig04} 
\end{figure}

\subsection{Local environment examples} \label{sec:subsample} 
\subsubsection{VFTS 385, 398, 470 and 581 } 
\begin{figure}
	\resizebox{\hsize}{!}{
	\includegraphics{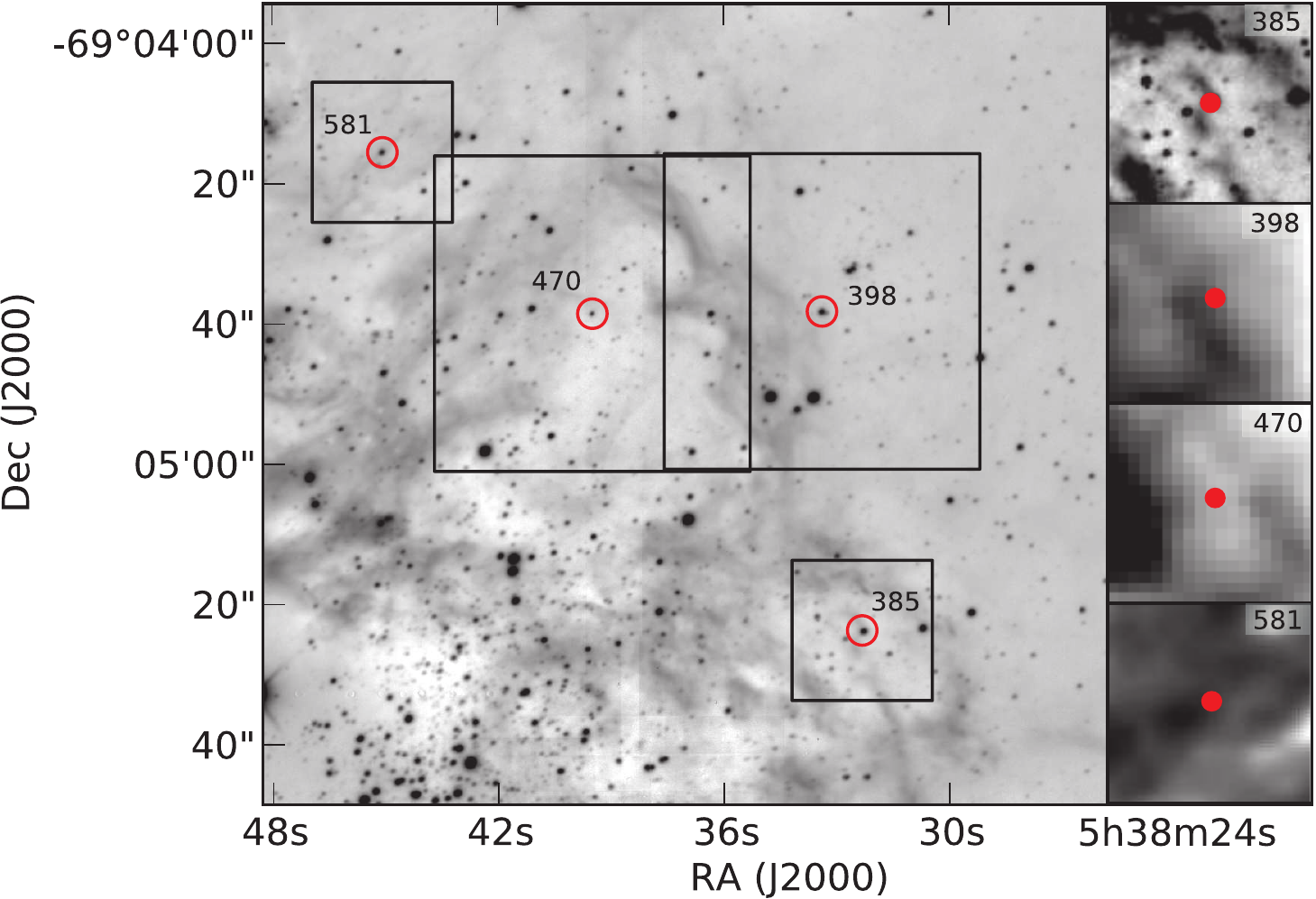}} \caption{Filamentary structures possibly associated with the four candidates: VFTS 385, 398, 470 and 581 (red circles). The background image is from the VLT HAWK-I K$_s$ band and the four subplots on the right are K$_s$, 70 $\mu$m, 70 $\mu$m and H$\alpha$/H$\beta$ \citep{Lazendic2003} maps for VFTS 385, 398, 470 and 581, respectively. The black boxes in the main image correspond to the field-of-view for the subplots on the right, where the 70 $\mu$m subplots are larger due to their lower resolution. Different stretches are used to highlight the filamentary associations in the subplots.} \label{fig:fig05} 
\end{figure}

From east-to-west in Fig.~\ref{fig:fig05} we have VFTS 581, 470, 398 and 385. We see that they are associated to filamentary structures from the VLT HAWK-I K$_s$ band, the \ext~extinction map \citep{Lazendic2003}, and the 70 $\mu$m map. The observation of VFTS 581 will be specifically modelled in Sect. \ref{sec:models} in order to see how visible its underlying cluster would be. VFTS 398 and 470 are both classified as Vz stars.

\subsubsection{VFTS 208 and 216} 
\begin{figure}
	\resizebox{\hsize}{!}{
	\includegraphics{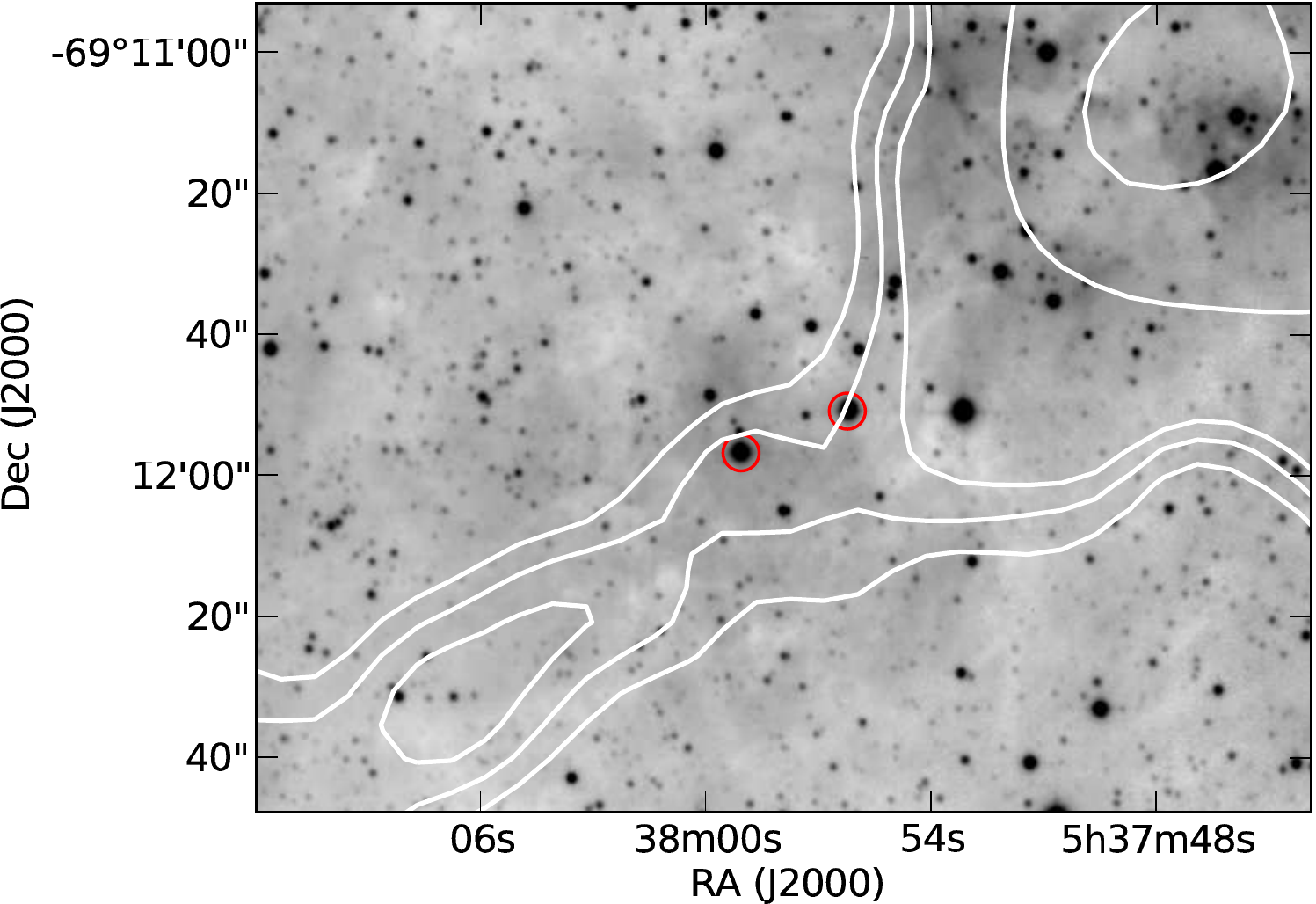}} \caption{$V$-band optical image with 70$\mu$m contours from {\it Spitzer} overlaid, in the region of VFTS 208 and 216. The stars are associated with a large filament extending southeast from the direction of NGC\,2060. The radial velocities of both stars agree to within 5 $\kms$. VFTS 216 may be related with a bow shock (see Fig.~\ref{fig:fig07}).} \label{fig:fig06} 
\end{figure}

\begin{figure}
	\resizebox{\hsize}{!}{
	\includegraphics{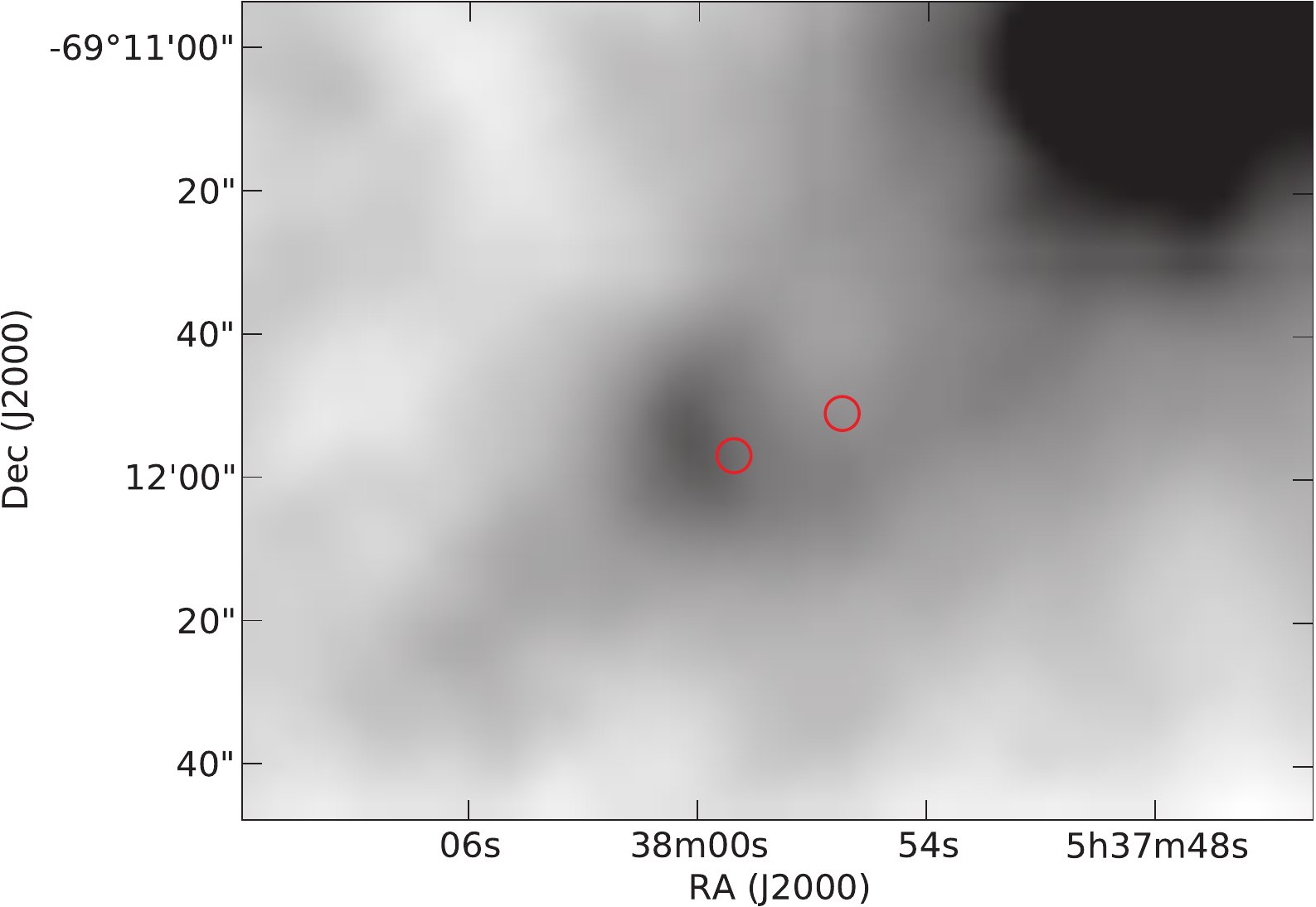}} \caption{VFTS 216 (left target) may possibly be related to a bow shock as seen above. The image is a 24 \micron~map from the SAGE Survey. If the O-star is indeed related to the bow shock, then its candidacy is ruled out as an isolated massive star. At the distance of the LMC, the bow shock is estimated to be 5 pc across.} \label{fig:fig07} 
\end{figure}

VFTS 208 and 216 are found to be associated with a large filamentary structure (Fig.~\ref{fig:fig06}) in the $70~\mu m$ Spitzer band, which is associated with NGC 2060, as seen in Fig.~\ref{fig:fig01}. No significant stellar populations are seen nearby, implying that they are relatively isolated. There may be a possible bow shock or bubble related to VFTS 216, which is shown in Fig. \ref{fig:fig07}. If there is a bow shock associated with the star, this would strongly suggest that it is a runaway, and did not form in isolation. The stellar surface distribution (based on HST V and I-band imaging) of VFTS 208 is shown in Fig.~\ref{fig:fig04}. We note that VFTS 208 is classified as an nfp star \citep{Walborn1973}. While the origins for the spectral peculiarities seen in such stars are still unknown (see \citealt{Walborn2010}), we retain the star in our discussion as it appears both relatively massive and young.

\subsubsection{VFTS 488 and 706}

\begin{figure}
	\resizebox{\hsize}{!}{
	\includegraphics{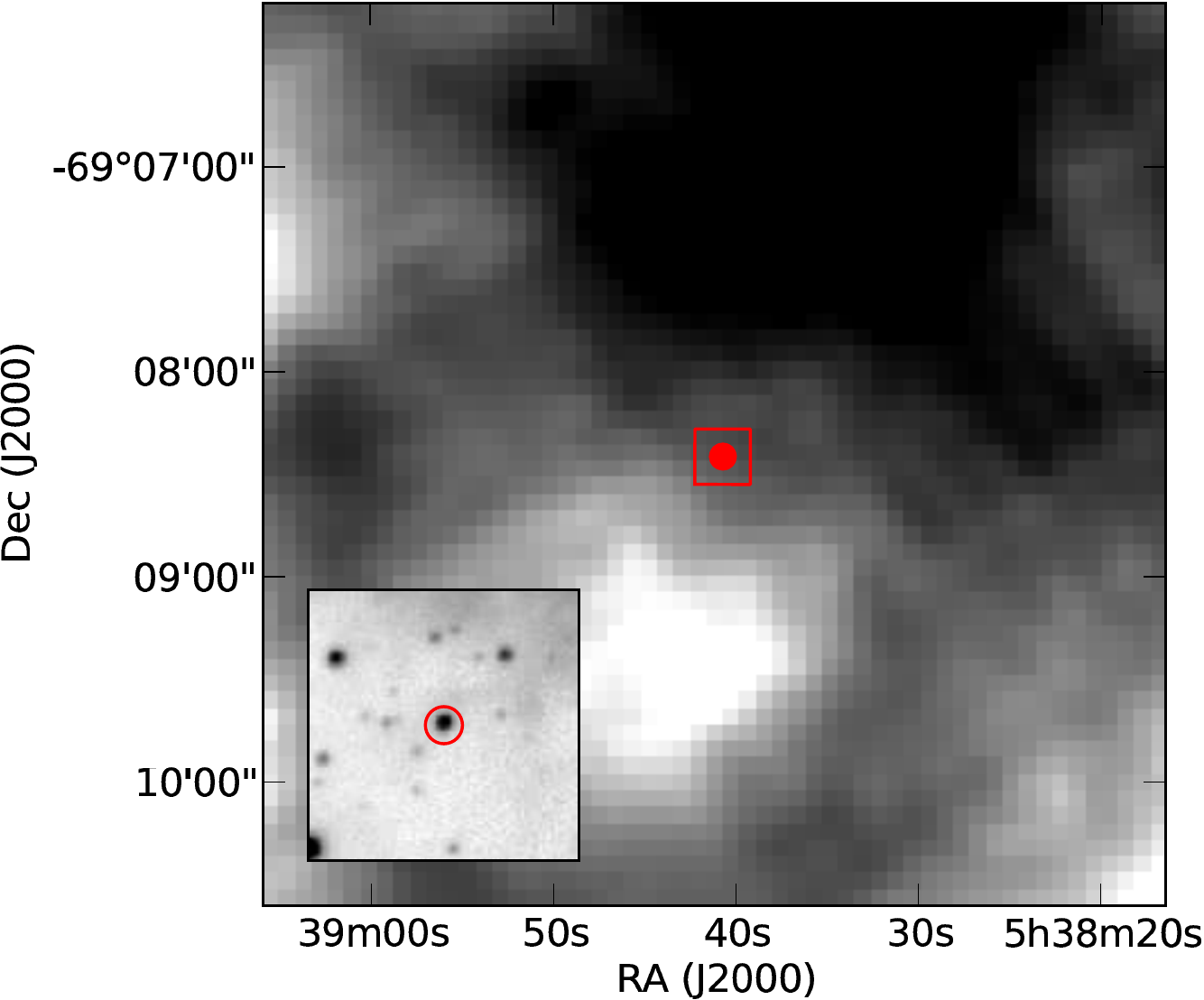}} \caption{VFTS 488 is shown in the 70 \micron~map as a red dot (main) and K$_s$ as a red circle (subset). The red box in the 70 \micron~map represents the subset image's field-of-view. VFTS 488 is associated with a filament in 70 \micron~and the K$_s$ subset image shows that it's located in a relatively sparse field of stars.} \label{fig:fig08} 
\end{figure}

VFTS 488 lies south of R136 at the edge of the dense 70 \micron~emission (see in Fig. \ref{fig:fig08}). It appears that VFTS 488 is associated with a filament, and both the VLT K$_s$ and HST F814W bands show its relative isolation from star clusters.

\begin{figure}
	\resizebox{\hsize}{!}{
	\includegraphics{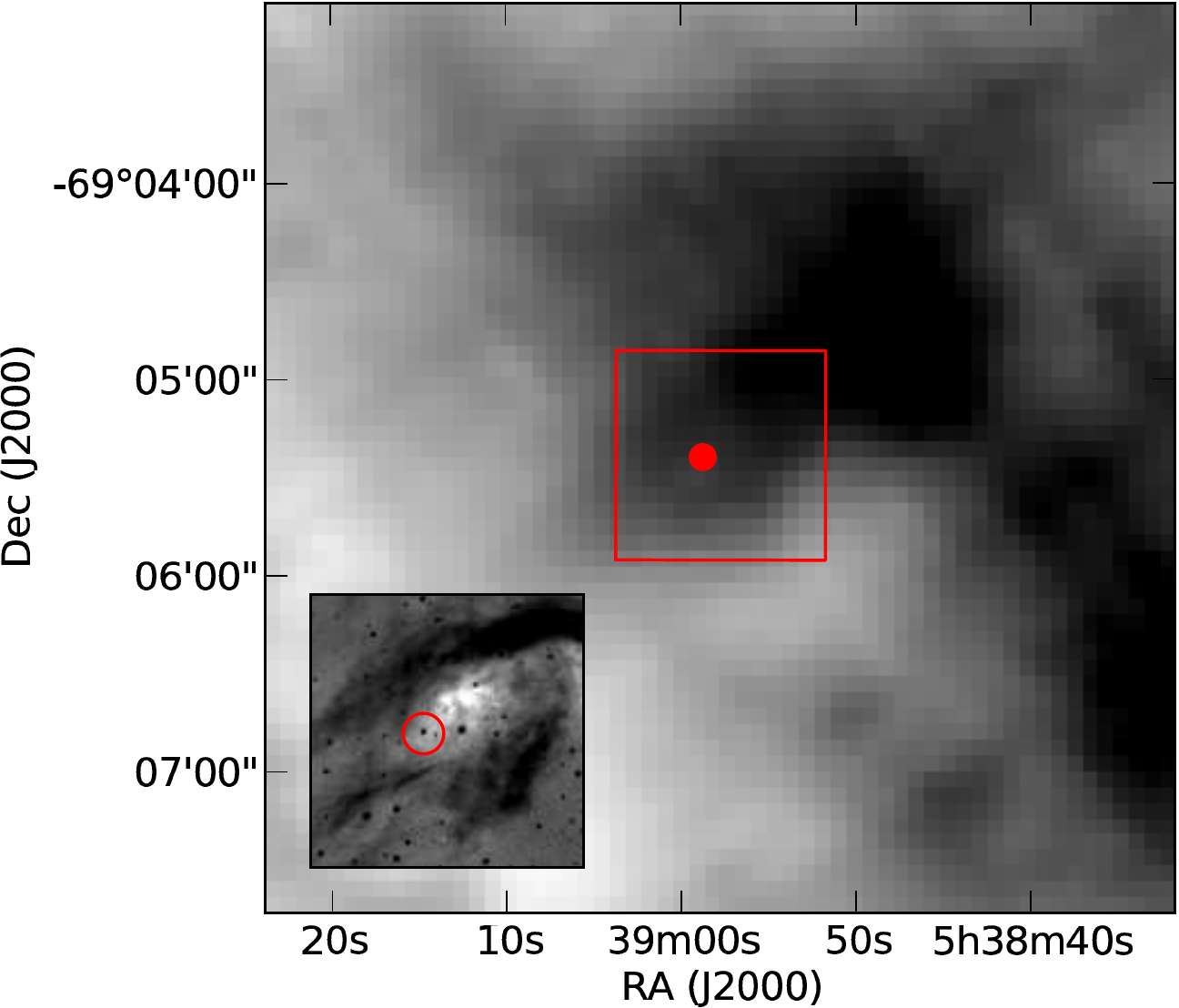}} \caption{VFTS 706 is shown in the 70 \micron~map (main) and K$_s$ (subset). The red box in the 70 \micron~map represents the subset image's field-of-view. The candidate is located in a region of high 70 \micron~emission, where filaments appear to be in the peripheral region. This is similarly seen in the subset where ionised material is near the candidate.} \label{fig:fig09} 
\end{figure}

Northeast of R136 is a prominent ionised filament structure that is seen both in the V and K$_s$ band. VFTS 706 is at the centre of the filament (see Fig. \ref{fig:fig09}). It can be seen in 70 \micron~that the region is associated cold dust. Both candidates are Vz stars.

\subsubsection{VFTS 682} \label{sec:682}

\begin{figure}
	\resizebox{\hsize}{!}{
	\includegraphics{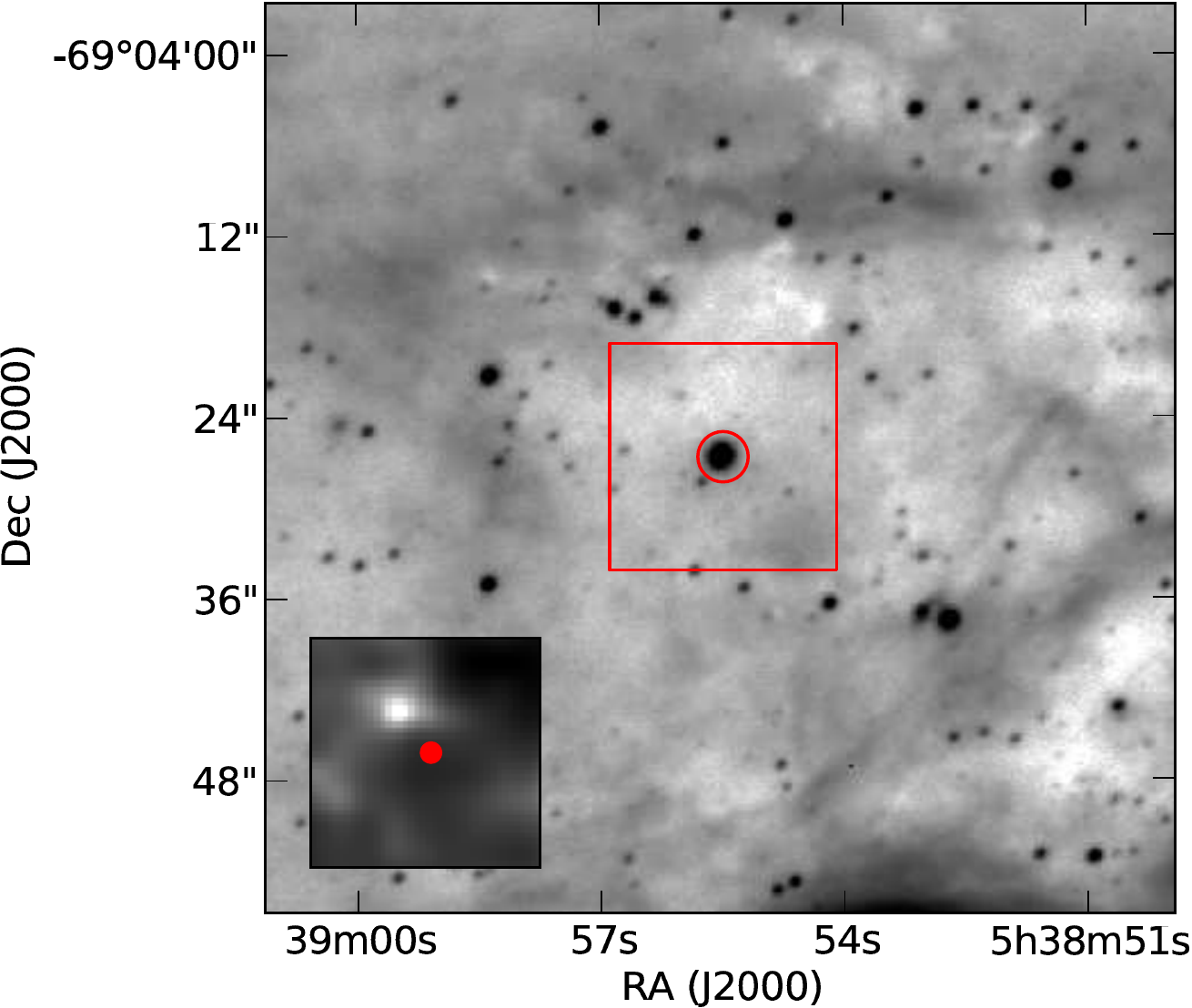}} \caption{VFTS 682, a Wolf-Rayet star that is discussed in further detail by Bestenlehner et al. (2011), is shown in the K$_s$ band (main) marked by a red circle and H$\alpha$/H$\beta$ derived map as a red dot (subset). The red box in the K$_s$ image represents the subset image's field-of-view. The candidate is located in a relatively sparse field of stars and the subset shows that's located in a region of relatively high A$_V$. VFTS 682 is the most massive candidate presented in this paper.} \label{fig:fig10} 
\end{figure}

Northeast of R136 we have VFTS 682, which does not fit our criteria as our best estimate of its radial velocity (from models of its emission lines) is outside the standard deviation from the mean. However, its young age and exceptionally high mass without evidence of clusters merits attention. Evans et al. (2011) discovered VFTS 682 to be a, previously unknown, Wolf--Rayet star (see Fig. \ref{fig:fig10}). They also noted it was within 4 pc of VFTS~702, a candidate young stellar object (YSO) from \cite{Gruendl2009}. The star displays excess emission in the mid-IR although the cause is still unclear, however the correlation with molecular gas suggests a region of ongoing star formation. VFTS 682 is the most massive star in our sample, with \cite{Bestenlehner2011} estimating an age ranging between 1 to 1.4 Myrs (i.e. in our youngest age range) from comparisons with evolutionary models. \cite{Bestenlehner2011} estimate that the A$_V > 4$ for the candidate which could make it difficult to detect lower mass stars around the candidate. However, according to the \mm\ relation, there should be a cluster of $\sim$ 3900~\msun~around it. This candidate is specifically modelled in Sect. \ref{sec:models}, and such a cluster would have easily been detected. \cite{Banerjee2012} proposes that VFTS 682 could be a {\it slow runaway}, but is only possible under their model assuming complete mass-segregation in the parenting cluster (R136) and that all massive stars are in binaries. Additionally, we find no evidence of bow shocks around the candidate at 24 \micron.

\subsection{Photometric completeness around the candidates} \label{sec:scm}

In order to test our photometric completeness as a function of magnitude and distance from each candidate source, we have estimated the 90\% completeness limit relative to the radius from the candidates. For this we used {\it PSFEx} \footnote[2]{PSFEx extracts models of the PSF from FITS images processed with SExtractor (http://www.astromatic.net/software/psfex). The software comes from the same team that maintains SExtractor} to generate artificial stars from modelled point spread functions (for each of the cameras used) and added them to the science frames (Fig. \ref{fig:fig02}). The artificial sources were given magnitudes between 16.5 and 24 mag, and 49 sources were added to the image in each iteration (in order to avoid excess crowding). For each magnitude step of 0.5, 50 iterations were carried out (resulting in 2450 sources per magnitude added). {\it SExtractor} \citep{Bertin1996} was used to detect the sources, and the resulting catalogues were used to test the success rate of detections near the candidates. An example of the procedure is given in Fig.~\ref{fig:fig03}, where we see (for VFTS 385, which is a typical source) we are 90\% complete at $V=23$~mag at 0.26~pc from the candidate O-star. This corresponds to a $\sim2$~\msun~star assuming an extinction estimate of $A_V = 1$. 

Additionally, the cumulative surface densities of sources as a function of radius away from each candidate was investigated using HST V and I-band imaging. An example is shown in Fig. \ref{fig:fig04}, which shows no evidence of stellar density increase around VFTS 208. For each source we adopted a conservative completeness limit (generally in the range of $V=$~21 to 23 mag) and focussed on radii greater than 1~pc. If a cluster was present, we would expect a rise in the number of stars per pc$^2$ from $\sim3$~pc and inwards (towards 0pc, the reference frame of the candidate). Similar results are seen for all the other candidates where no stellar density increase is observed. The information from this section will be used further in Sect. \ref{sec:models} to explore the constraints on any underlying cluster that may be present.

Upon visual inspection of higher resolution HST data from the WFC3 \citep{DeMarchi2011} of VFTS 385, 392, 577 and 706 we see no evidence of clustering in the F555W and F814W filters. We do not present the data itself in this paper.

\subsection{Binary detection probability} \label{sec:binarity}

To estimate the fraction of undetected spectroscopic binaries in our sample, we have quantified the observational biases of each candidate using the Monte Carlo methods from \cite{Sana2009}. We estimated the probability (P$_{\textrm{ detect}}$) to detect binarity in each candidate, adopting a primary mass on the basis of their spectral types and with binary properties randomly drawn from uniform cumulative distribution functions (CDFs) of mass-ratios and eccentricities, and a bi-modal CDF in $\log$\,P (as described by \citealt{Sana2011}). Assuming a random orientation of the system in space and a random time of periastron passage, we then apply the specific observational sampling of each object, adding noise to the RV signal that corresponds to the error measurements at each epoch. Finally, we consider whether the simulated system would have been detected if the amplitude of the RV signal is larger than an adopted detection threshold. 

For all the O-star candidates, except VFTS 706, we adopt a threshold of 20 $\kms$ from the mean. These levels are chosen to ensure no false detections at the 99.99\% confidence level ($>$3$\sigma$). VFTS 706 shows evidence of faster rotation than the others, so we have fixed the threshold to 35 $\kms$, yielding no false detections at a 99.7\% confidence. Table \ref{tab:binarity} provides the computed detection probabilities across three ranges of periods as well as across the full period range. This shows that it is unlikely that our sample contains undetected short and intermediate period spectroscopic binaries. For period larger than one year, our detection probabilities become relatively small. Under the adopted hypothesis, {\it the average detection probability is $0.83 \pm 0.05$ up to a period of $\sim$8.6 yr}. While 10 to 20\% of the objects in our sample might still be undetected, likely long-period spectroscopic binaries, these simulations allow us to conclude that most of our targets are either single stars or very wide/large mass ratio pairs so that the measured systemic velocity is left unaffected by the companion. Additionally, the spectra were inspected for SB2 systems.
\begin{table}
	\caption{Average binary detection rates.} \centering 
	\begin{tabular}
		{lcccc}VFTS ID&2$^d$-10$^d$&10$^d$-365$^d$&365$^d$-3160$^d$&2$^d$-3160$^d$\\
		\hline 089&0.989&0.897&0.424&0.852\\
		123&0.990&0.886&0.430&0.855\\
		208&0.993&0.892&0.456&0.862\\
		216&0.995&0.899&0.460&0.864\\
		382&0.990&0.917&0.490&0.874\\
		385&0.995&0.897&0.449&0.861\\
		392&0.992&0.886&0.425&0.851\\
		398&0.993&0.882&0.417&0.854\\
		470&0.994&0.919&0.485&0.873\\
		488&0.993&0.868&0.411&0.848\\
		537&0.996&0.905&0.473&0.867\\
		577&0.993&0.888&0.431&0.854\\
		581&0.996&0.905&0.459&0.870\\
		706&0.976&0.783&0.266&0.786\\
		849&0.992&0.923&0.456&0.873\\
		\hline 
	\end{tabular}
	\begin{flushleft}
		{{\bf Notes.} Average detection rate, P$_{\textrm{detect}}$ from the simulation of 10000 samples of 15 stars: $0.83 \pm 0.05$. The WN5h (Wolf-Rayet) star, VFTS 682, is not included in this, as it is not a standard O-type star.} 
	\end{flushleft}
	\label{tab:binarity} 
\end{table}

\section{Discussion} \label{sec:conclusions}

\subsection{Modelling potential underlying clusters} \label{sec:models}

\subsubsection{Visual examples of cluster presence} In order to check if we are sensitive to clusters that may be associated with the candidates according to the \citet{Weidner2010} \mm\ relation, we use the {\it MASSCLEAN} package \citep{Popescu2009} to generate illustrative examples. The package generates images (at a given distance and resolution), of simulated clusters with stars drawn from a Kroupa IMF \citep{Kroupa2001} and spatially distributed according to a King profile \citep{King1962}. For the analysis we adopt King profile radii similar to those reported by \cite{Werchan2011} for the modelled clusters. Stellar masses are drawn such that the \citet{Weidner2010} \mm\ relationship is recovered from these simulations. All simulations were carried out adopting the resolution of the ACS WFC.

\begin{figure*}
	\centering 
	\includegraphics[width=17cm]{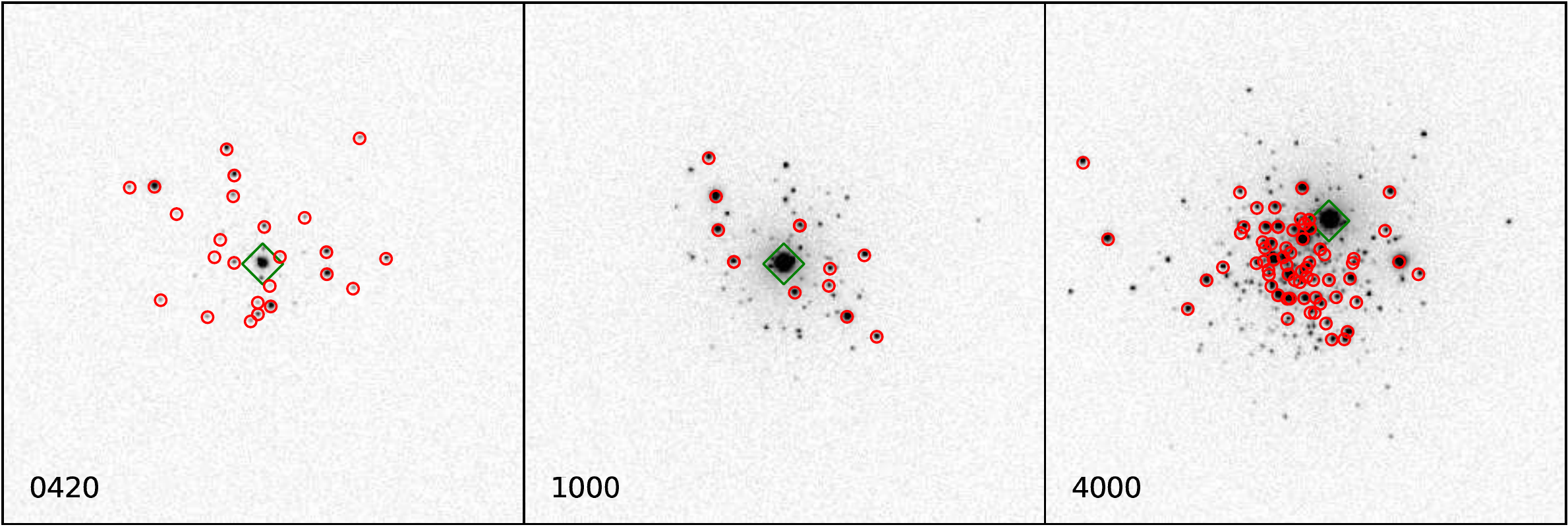} \caption{From left to right, the 5$\times$5 pc (20''$\times$20'') images of the simulated {\it MASSCLEAN} clusters of 420, 1000, and 4000 \msun. The images were made of simulated clusters, with HST ACS resolution, at the distance of the LMC in the V-band. The green diamonds mark the brightest/most massive star in the cluster. The red circles indicate stars that would have been detected on the actual images (see Fig.~\ref{fig:fig03}). In all cases, the underlying cluster would have been detected.} \label{fig:fig11} 
\end{figure*}

We generated three clusters, whose most massive stars were 25, 45, and 100~\msun. The 25 \msun~star represents the lower mass estimates of the candidates given in Table \ref{tab:diffmass}, while the 45 and 100 \msun~stars represent VFTS 581 and 682 (Table \ref{tab:tab01}), respectively. The total cluster masses are 420, 1000 and 4000~\msun, respectively, and are shown in Fig.~\ref{fig:fig11}. In order to account for the effects of extinction (which we assume affect all cluster members equally), we then scaled the images such that the most massive star had the same V-band apparent magnitude as the candidate which we were modelling. 

In Fig.~\ref{fig:fig11} the most massive star is indicated by a green diamond. Note that in the 4000~\msun\ image, the most massive star is not in the centre of the cluster, this is simply due to the fact that the positions of each of the stars is distributed stochastically, regardless of mass. We then applied the completeness curve from Fig.~\ref{fig:fig03}, assuming that all stars above the 90\% completeness limit (at a given radius) would have been detected. All stars that pass this criteria are circled (red) in the images. In the middle panel of Fig.~\ref{fig:fig11} fewer stars would have been detected than in the left (420~\msun\ case) panel. This is due to the difference in reddening between the two model stars (for similarity to the candidates).

In all cases we would likely have observed an underlying cluster, which is evident when comparing Fig. \ref{fig:fig02} and Fig. \ref{fig:fig11}, suggesting that these stars are isolated compared to what one would expect for star formation drawn from sorted sampling. Each of the O-star candidates should have B-stars present and several should have other O-stars in their associated cluster.

\subsubsection{Monte Carlo simulations} \label{sec:mcs} Alongside the visual examples of the typical clusters that should be present around the massive O-star candidates, we conduct three comprehensive sets of Monte Carlo simulations.The exact same parameters of the most massive stars and clusters given in the previous section are adopted, but with uniform dispersion of cluster masses for each simulation set (SS\#). The cluster mass ranges are extrapolated from the $1/5$ and $5/6$ quantile dispersions reported from Fig. 5 in \cite{Weidner2010}. Details of the SS\# are listed below. 
\begin{enumerate}
	
	\item {\bf SS1:} 10,000 runs with median cluster mass of 420 \msun with its most massive star at 25 \msun. The cluster mass dispersion range is $10^2 - 10^3$ \msun.
	
	\item {\bf SS2:} 10,000 runs with median cluster mass of 1000 \msun with its most massive star at 40 \msun. The cluster mass dispersion range is $10^{2.2} - 10^{3.6}$ \msun.
	
	\item {\bf SS3:} 10,000 runs with median cluster mass of 4000 \msun with its most massive star at 1000 \msun. The cluster mass dispersion range is $10^{3.1} - 10^{4.1}$ \msun. 
\end{enumerate}

\begin{figure*}
	\centering 
	\begin{tabular}
		{lll} 
		\includegraphics[width=5.7cm, angle=0]{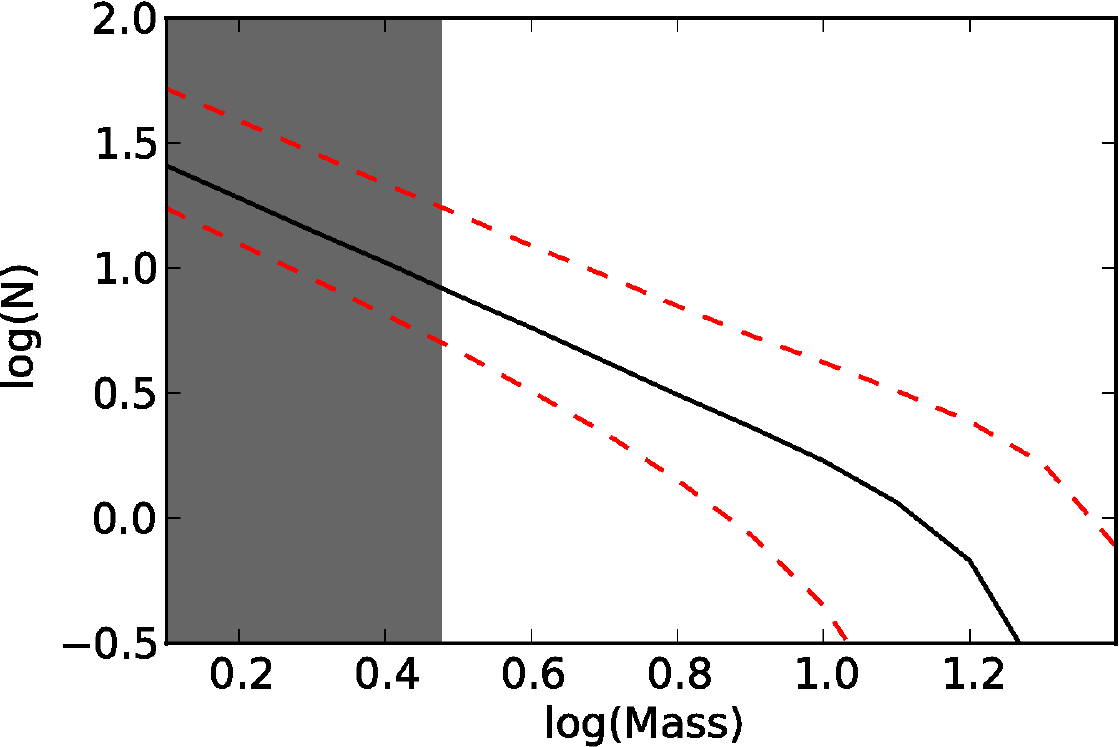} & 
		\includegraphics[width=5.7cm, angle=0]{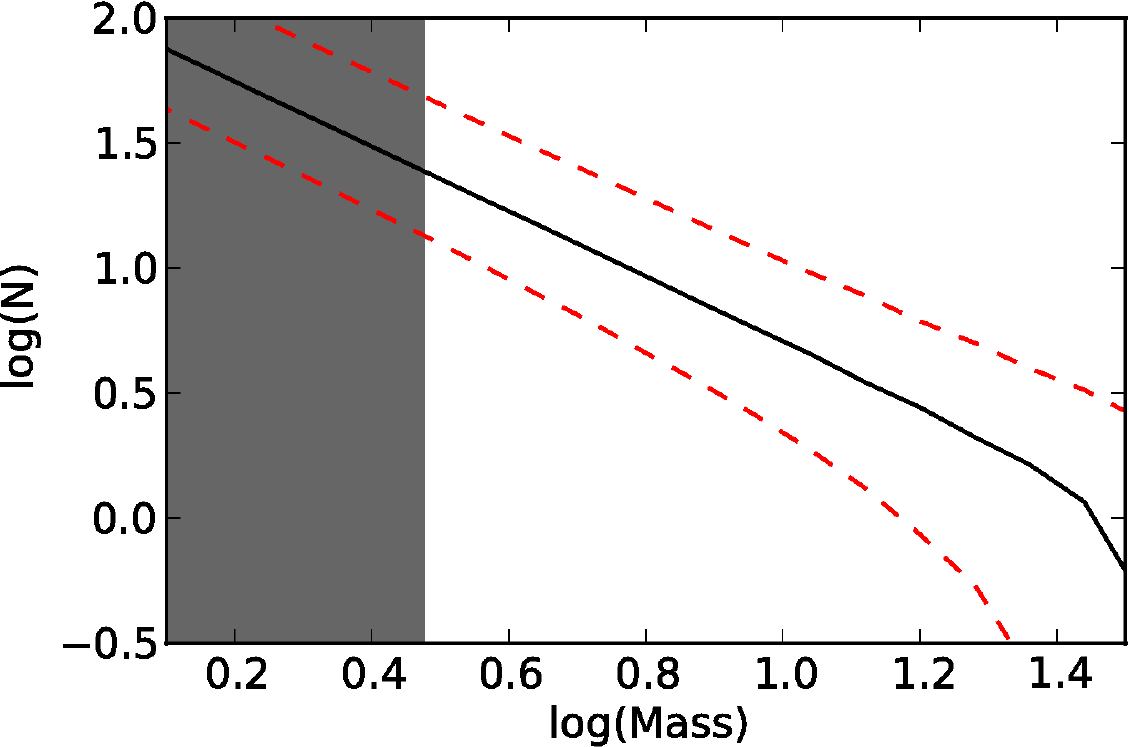} & 
		\includegraphics[width=5.7cm, angle=0]{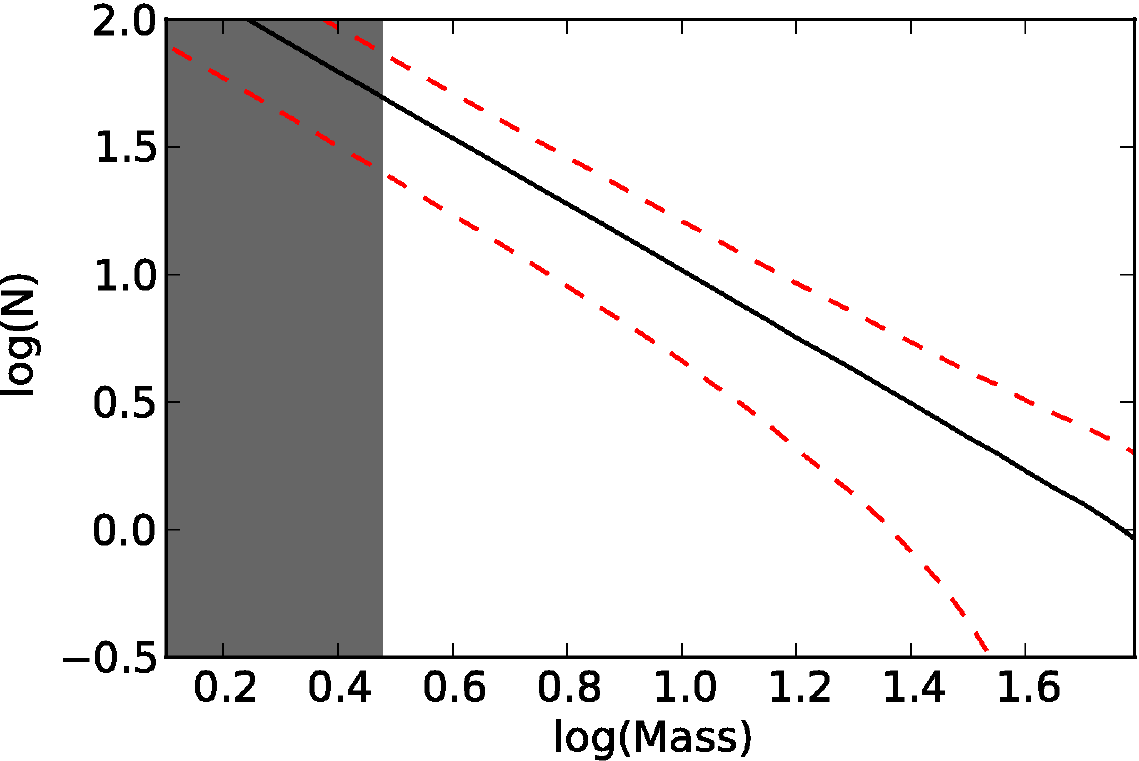} \\
	\end{tabular}
	\caption{{\it[From left to right]} The expected number of observable stars associated with a 25, 40, and 100 \msun\ massive star in 420, 1000, and 4000 \msun\ clusters, respectively based on the 30,000 Monte Carlo simulation runs. The black solid lines are the mean number of stars per mass bin and the dashed red lines are the 1$\sigma$ dispersion. Stars with masses below 3 \msun\ are greyed out on the left hand side of the plot, as those to the right of it are observable. The expected number of observable stars with a 1$\sigma$ dispersion from the mean are 11, 53, and 148 assuming $A_V \sim 1$. According to the estimated number of excess stars for the candidates (see Table \ref{tab:clusters}) we should have detected such cluster presence around the 25 \msun\ candidates at minimum.} \label{fig:fig12} 
\end{figure*}

To determine the observable number of stars in the simulations we remove stars below our estimated mass sensitivity limit of 3 \msun. Furthermore, we assign cluster positions to each of the stars using the same King profile parameters as mentioned in the previous section and remove any stars within 0.15 pc of the most massive star in the cluster. This reflects the completeness limit as a function of radius discussed in Sect. \ref{sec:scm}. In Fig. \ref{fig:fig12} the number of observable stars with a 1$\sigma$ dispersion is shown. For the cases of SS1, SS2 and SS3 we observe at minimum 11, 53, and 148 numbers of stars at the 1$\sigma$ dispersion level from the mean assuming $A_V \sim 1$. 

If we place a system like the Trapezium cluster (just the inner $0.1$~pc of the Orion Nebula Cluster) in the LMC we would not be able to resolve it. However, with an effective radius of $2$~pc (e.g. \citealt{PortegiesZwart2010}) we would have readily detected the surrounding cluster with minimum of $\sim40$ excess stars in a 5 pc radius. This is particularly relevant for the \mm~relationship presented by \citet{Weidner2010}, for which the full Orion Nebula Cluster must be used for the most massive star, $\theta^1 \mathrm{Ori~C}$ ($48$~\msun, \citealt{Kraus2007}) to fit the relation.

\subsection{Comparing Monte Carlo simulations to observations}

If the \mm\ relationship is correct we should have an excess number of stars within a 5 pc radius of each of the candidates. Around each candidate we measured the number of observable stars above 3 \msun\ in the HST data and compared them to the Monte Carlo simulation sets discussed in the previous section. Two apertures were used to estimate the number of excess stars relative to the background and foreground stars. An inner aperture of 5 pc and an field aperture of 10 pc, which are denoted as NN$_{\textrm{inner}}$ and NN$_{\textrm{field}}$, respectively. We define the number of excess stars (field star corrected) in the inner aperture as N$_{\textrm{excess}} = (\mbox{NN}_{\textrm{inner}} - \mbox{NN}_{\textrm{field}}) \times \pi~r^2$ where $r = 5$ pc. Upon comparing the number of excess stars to the number of expected stars if a cluster is present from the simulations (see Table \ref{tab:clusters}) we see no evidence for clusters around the candidates.

\begin{table}
	\caption{Number of stars around each candidate based on observations and Monte Carlo stellar cluster simulations.} \centering 
	\begin{tabular}
		{lcccc} VFTS & NN$_{\textrm{inner}}$ & NN$_{\textrm{field}}$ & N$_{\textrm{excess}}$& N$_{\textrm{sim}}$\\
		\noalign{\smallskip}& [stars pc$^{-2}$] & [stars pc$^{-2}$] & [star count] & [star count]\\
		\noalign{\smallskip}\hline 089& 0.09& 0.07& 2& 53\\
		123& 2.01& 2.23& $-$17& 53\\
		208& 2.93& 2.61& 25& 53\\
		216& 2.67& 2.51& 13& 53\\
		382& 0.32& 0.22& 7& 53\\
		385& 0.32& 0.38& $-$5& 53\\
		392& 0.35& 0.22& 11& 53\\
		398& 0.37& 0.46& $-$7& 53\\
		470& 0.51& 0.48& 3& 53\\
		488& 0.64& 0.39& 19& 53\\
		537& 0.22& 0.29& $-$6& 53\\
		577& 0.70& 0.66& 3& 53\\
		581& 0.31& 0.41& $-$9& 53\\
		682& 0.05& 0.07& $-$2& 148\\
		706& 0.22& 0.12& 8& 53\\
		849& 0.28& 0.19& 7& 11\\
		\hline 
	\end{tabular}
	\begin{flushleft}
		{{\bf Notes.} If the \mm\ relationship is correct we should see an excess number of stars within a 5 pc radius of each of the candidates. Around each candidate we measured the number of observable stars above 3 \msun\ from the HST images and compared the number of excess sources (N$_{\textrm{excess}}$) to the Monte Carlo simulations (N$_{\textrm{sim}}$). This shows no evidence of clustering around the massive star candidates.} 
	\end{flushleft}
	\label{tab:clusters} 
\end{table}

\subsection{Mass and age discrepancies} \label{sec:diffmass}

We use the spectral types of our candidates to derive their age and masses via \cite{Weidner2010c} as shown in Table \ref{tab:tab01}, but these properties can also be obtained from other methods. To assess potential uncertainties we have also obtained physical parameters using the Galactic spectral type-$T_{\textrm{ eff}}$ calibration of \cite{Martins2005}. Absolute visual magnitudes were obtained by correcting the photometry from Evans et al. (2011) for the effects of extinction, with stellar luminosities estimated from the bolometric correction calibration of \cite{Martins2005}. Masses and ages follow from comparison with non-rotating, LMC metallicity stellar evolutionary models and isochrones \citep{Meynet1994, Lejeune1997}.

We acknowledge that the use of an alternative spectral type-$T_{\textrm{eff}}$ calibration (e.g. \citealt{Massey2009}) may lead to increased stellar luminosities/masses. Conversely, decreased stellar masses would be inferred from contemporary evolutionary models allowing for rotation \citep{Maeder2001,Brott2011}. Results are presented in Table \ref{tab:diffmass}, although refined parameters await detailed analyses, which are currently in progress. Individual stellar masses are on average $<$25\% below those derived using spectral types \citep{Weidner2010c}. Nevertheless, our key conclusions are unaffected by the method used to derive masses since the parent clusters, if present, would still be observable. Except for VFTS 682, all candidates (including VFTS 208) would meet our criterion and be assigned grade 1 or 2 according their estimated ages under this method.
\begin{table}
	\caption{Comparison of candidate masses using both evolutionary models with isochrones and \cite{Weidner2010c} models.} \centering 
	\begin{tabular}
		{lccccc} VFTS & $\log$(T$_{\textrm{eff}}$) & $\log$(L) & Age & Mass & Mass$_{\rm WV10}$\\
		\noalign{\smallskip}& [K] &[$L_{\odot}$]& [Myr] &[$M_{\odot}$]&[$M_{\odot}$]\\
        \noalign{\smallskip}\hline 089 & 4.58 & 4.91 & 1.5 & 25 & 33\\
		123 & 4.58 & 4.91 & 1.5 & 25 & 33\\
		208 & 4.57 & 5.62 & 3.5 & 43 & 46\\
		216 & 4.63 & 5.79 & 2.2 & 57 & 53\\
		382 & 4.62 & 5.04 & 0.0 & 29 & 48\\
		385 & 4.62 & 5.36 & 1.8 & 37 & 48\\
		392 & 4.58 & 4.82 & 0.0 & 24 & 33\\
		398 & 4.60 & 5.36 & 2.5 & 36 & 36\\
		470 & 4.58 & 4.81 & 0.0 & 24 & 44\\
		488 & 4.59 & 5.10 & 2.0 & 29 & 36\\
		537 & 4.61 & 4.88 & 0.0 & 25 & 44\\
		577 & 4.59 & 4.97 & 1.0 & 27 & 36\\
		581 & 4.62 & 5.01 & 0.0 & 29 & 48\\
		706 & 4.58 & 4.96 & 2.0 & 26 & 33\\
		849 & 4.57 & 4.91 & 2.5 & 24 & 30\\
		\hline 
	\end{tabular}
	\begin{flushleft}
		{{\bf Notes.} Derived properties of our isolated candidates (O-stars) from comparisons with evolutionary models and isochrones with a metallicity appropriate to that of the LMC \citep{Meynet1994, Lejeune1997}. Although the masses differ to those in the last column from \cite{Weidner2010c}, our key conclusions are unaffected.} 
	\end{flushleft}
	\label{tab:diffmass} 
\end{table}

\subsection{Filamentary structures in 30 Doradus} \label{sec:filaments} The 30 Doradus region is largely affected by R136. The gaseous filaments that we consider in method four for associating with the O-stars (see Sect. \ref{sec:method}) could possibly be problematic for the slightly older stellar population. In particular, the grade 3 candidate, VFTS 208, since the filaments that such a star formed in, could have been moved/destroyed by feedback from the nearby R136 cluster (e.g. \citealt{Tenorio-Tagle2006}).

However, others argue that since the ISM is filamentary and heterogeneous (e.g. \citealt{Andre2010,Bergin2007}), the effects of ionising sources on the gas will be less significant than what \citealt{Tenorio-Tagle2006} show in their simulations (e.g. \citealt{Dale2011}).

Note that as an additional constraint on the association between stars and gas, we measured line-of-sight gas velocities from the [NII]6583, [SII]6717, [SII]6731, and H$\alpha$ nebular emission lines superimposed on the stellar spectrum (see Table \ref{tab:tab01}). When multiple gas velocity components are identified, only the one that is closest to RV$_{\textrm{star}}$ is presented. We can see that 7 of the 16 candidates have a radial velocity for the ionised gas (RV$_{\textrm{ISM}}$) within $\sim$5~$\kms$ of the mean RV$_{\textrm{star}}$, while the agreement between RV$_{\textrm{ISM}}$ and RV$_{\textrm{star}}$ is within 15~$\kms$ for most candidates.

\subsection{Bow-shocks and the ISM in 30 Doradus} \label{sec:bowshocks} \cite{Gvaramadze2010} discovered bow shocks around two isolated massive stars a few hundred pc away from R136. They argue that these OB stars are consistent with being ejected from the stellar cluster. Their discovery proves that bow shocks can form in the 30 Doradus region (or at least in its surroundings) and that these structures can be detected in Spitzer 24 $\mu$m images. So, while the presence of a bow shock suggests the runaway nature of an isolated OB star \citep{Gvaramadze2010}, what can one conclude for the isolated candidates presented in this paper that do not show evidence of bow shocks? We can estimate the minimum required runaway velocity to explain the apparent isolation of the candidates considered in this paper. From the observed projected distance $d$ from R136 (14 to 130 pc ) and assuming an age $t \leq 1$Myr 
\begin{displaymath}
	v \ge 9.8~\bigg(\frac{d}{{10 \rm ~pc}}\bigg) \, \bigg(\frac{t}{{\rm Myr}}\bigg)^{-1} \, ~\kms 
\end{displaymath}
we obtain minimum runaway velocities between 14 and 130 $\kms$. These velocities are higher than the isothermal sound speed in a warm ISM. In the case of 30 Doradus the average gas temperature is estimated to be around $10^4$ K, with radiation pressure appearing to be negligible outside 10 pc from the central cluster R136 \citep{Pellegrini2011,Peimbert2003}. This results in an isothermal sound speed of order 10 $\kms$. Inside the cavities containing hot, low density gas, the sound speed can increase up to 100 $\kms$. However such cavities are mostly observed in the inner region of 30 Doradus \citep{Pellegrini2011}. Therefore, if our isolated massive stars are runaways, they are likely to be moving supersonically with respect to the local ISM.

In a fraction of supersonically moving runaway OB stars, the interaction between the stellar wind and the ISM will produce observable bow shocks (e.g. \citealt{Comeron1998}). This is supported by observations of such objects, both in the Galaxy (e.g. \citealt{vanBuren1995,Huthoff2002, Gvaramadze2008}) and in the Magellanic Clouds \citep{Gvaramadze2010,Gvaramadze2011}.

The absence of bow shocks in a large sample of isolated OB stars could hint to the fact that not all of them are runaways. \cite{Gvaramadze2010} found that $\sim30\%$ of the runaways in their sample in the LMC had bow shocks. This is consistent with Galactic studies of bow shocks as well where $20-40\%$ of well established runaway OB stars has a detectable bow shock \citep{vanBuren1995,Huthoff2002}. {\it Hence, for the LMC we would expect $\sim5$ of the 16 candidates to have bow shocks if they are all runaways.} While this argument is clearly not conclusive, in line with recent efforts of \cite{Gvaramadze2010,Gvaramadze2011}, we suggest that the investigation of the presence/absence of bow shocks around large enough samples of isolated O-stars could be an important way to constrain their formation mechanism.

\subsection{Number of isolated stars} \label{sec:number} What fraction of stars would we expect to form in isolation, generally? Not only does the answer depend on how the IMF is formed (e.g. sorted or stochastic sampling) but also on the intrinsic spatial distribution of stars at the time of their formation, and how this probability function is sampled. Historically, if one {\it assumes} that all stars are formed in ``clusters'', then clusters are the basic unit of star formation, from which stars are then sampled. However, if ``clusters'' in fact do not represent a basic unit of star formation, but instead there exists a distribution of surface densities (where ``clusters'' merely represent the high-surface density tail of the distribution) then one needs to know how the spatial distribution is sampled. Thus, for each IMF sampling algorithm, one would first need to draw the mass of the star, and then the surface density distribution in order to make a stellar population. At the moment, it is unclear how the surface or volume density distribution of young stars depends on environment (see \citealt{Bressert2010}), hence we cannot say what fraction of isolated stars would be expected, within this scenario.

\section{Summary and implications} \label{sec:summary}

We have presented a new method to identify massive stars that formed in relative isolation. Applying the method to the 30~Doradus region in the LMC, 15 O-star candidates are found, where 11 are Vz stars, that may have formed in isolation. The method uses precise radial velocities of the stars to rule out massive binaries and runaways along the line of sight. Additionally, high resolution imaging is used to constrain the possible presence of a stellar cluster around the stars. Using extensive Monte Carlo stellar cluster simulations we confirm that our observations are sufficiently sensitive to rule out typical stellar clusters that should be associated with the candidates if the \mm\ relation is correct. Finally, we search for gas/dust filaments that are associated with the massive stars (at least in projection), i.e., the birth-sites of stars, to mitigate the possibility that the stars are runaways in the plane of the sky. The gaseous filamentary structures were identified using a heterogeneous set of techniques, wavelengths and instruments. This could be significantly improved through the use of high-resolution mid-IR imaging of the cold dust in the region, such as that provided by the {\it Herschel Space Telescope}. The observations presented here can constrain theories of massive star formation as well as scenarios for sampling from the stellar IMF.

All the candidates, except VFTS 682, have met the criteria presented in Sect. \ref{sec:method}. We have mentioned several caveats, including filamentary structures and stellar age issues and possible bow shock association that could be problematic for grade 2 and 3 candidates. Hence, the best candidates of the 16 that withstand the caveats are the 11 Vz (ZAMS) stars, where their likely young ages, high mass, robustness to different stellar model parameters, and general close agreement between RV$_{\textrm{star}}$ and RV$_{\textrm{ISM}}$ is within 15 \kms, with the exception of 682 (40 \kms) for which formal uncertainties in stellar RV measurement are likely too low.

Isolated massive star formation is possible according to the monolithic formation scenario (e.g., \citealt{Krumholz2009}), whereas the competitive accretion scenario implies that massive stars can only form in sufficiently massive clusters (e.g., \citealt{Bonnell2004}). The observations presented here suggest that competitive accretion might not be the only mechanism responsible for the formation of massive stars. Whether a massive-star forms in a cluster or not can be affected by an external agent, such as triggering, but once star formation starts in the cluster radius of the massive star, external triggering is no longer an important factor \citep[see ][]{Dale2011}.

We have found 16 candidates in the LMC that most likely formed in isolation or in sparse clusters that do not follow the \mm~as reported by \cite{Weidner2010}. This supports the conclusions of \citet{Lamb2010} who studied O-stars in the SMC which had sparse clusters associated with them. Both studies attempt to remove the possibility that the O-stars investigated are runaways from a nearby cluster. \citet{Chu2008} found that 4\% of massive YSO candidates (with some contaminant, \citealt{Gruendl2009,Evans2011}), in the LMC were located far enough away from young clusters and/or associations that they most likely formed in isolation. Additionally, lower in the stellar mass range, a collection of Ae/Be stars have recently been shown to not originate in clusters \citep{WheelWright2010a}. Finally, \cite{Eldridge2012} notes that $\gamma$-Velorum is a $\sim65$\msun\ WR+O binary system in a cluster with a total mass of $250-350$~\msun, which is well below that expected from the \mm\ relation.

An additional constraint that not all of the surrounding O-stars in 30~Doradus are runaways, can be obtained by comparing R136 and the Galactic young massive cluster NGC~3603. The clusters have similar ages, masses, and densities, yet NGC~3603 does not appear to have any massive stars outside the central $\sim1$~pc \citep{moffat94}. Since the number of runaways scales with the central cluster density, NGC~3603 and R136 are expected to have similar numbers. Hence, the observed differences are likely a reflection of the formation of the cluster and surrounding environment.

Hence, the observations presented here (and those cited above) appear to contradict the \mm\ relation shown in \citet{Weidner2010}, in that large clusters are not necessary to form massive stars. This favours a scenario that the stellar IMF is sampled stochastically. Amongst the candidates, we have shown that short-period binaries are very rare. There may be unresolved lower-mass stars and a handful of long-term binaries that accompany the candidates, however, if present these stars wouldn't be enough to raise the \mm\ to fit the sorted sampling scenario. The fact that the \mm\ relation does not appear to be causal, but rather statistical, raises questions regarding the Integrated Galactic IMF (IGIMF) scenario \citep{Weidner2006}, where galaxies that form only low mass stellar clusters will never form high-mass stars. This has many potentially large implications for extragalactic observations (c.f., \citealt{Bastian2010}), especially for low-mass galaxies.

The observation of the formation of high-mass stars in isolation, along with the age spread within the 30~Doradus region (e.g., \citealt{Walborn1997}), is consistent with a hierarchical distribution of star formation, in time and space (e.g. \citealt{Efremov1998}, \citealt{Bastian2009}). This means that star formation does not happen in a quantised way on the group level, i.e., not all stars form in compact coherent structures like clusters. The spatial distribution of star formation appears to be a continuous distribution in surface density \citep{Bressert2010} of which the stars discussed here would fall into the extreme low surface density end of the distribution. Additionally, clusters are sensitive to how one exactly defines them, as the distribution of clusters and associations seem so overlap at young ages \citep{Gieles2011}.

\section*{Acknowledgments}

We would like to thank Nolan Walborn, Sally Oey, John Bally, Leonardo Testi, and Adam Ginsburg for their comments and suggestions. Additionally, we thank Jasmina Lazendic-Galloway for sharing \ext~extinction map with us and Carsten Weidner for interesting discussions. Based on observations at the European Southern Observatory Very Large Telescope in programme 182.D-0222. Based on observations made with the NASA/ESA Hubble Space Telescope, and obtained from the Hubble Legacy Archive, which is a collaboration between the Space Telescope Science Institute (STScI/NASA), the Space Telescope European Coordinating Facility (ST-ECF/ESA) and the Canadian Astronomy Data Centre (CADC/NRC/CSA).

\appendix

\section{Hubble archival data} \label{sec:hst} We make use of HST images that were retrieved through the Hubble Legacy Archive web service. The images came from a collection proposals and programs. A table is provided below to summarise the different data and their properties which are important for this paper. 
\begin{table}
	\caption{The details of the HST data used in this paper.} \centering 
	\begin{tabular}
		{lccc} VFTS& HST ID& Filter& Integration Time\\
		& & & [s]\\
		\hline 089& APPP-LMC& F814W& 3200\\
		123& 9741& F814W& 3200\\
		208& 9741& F814W& 2800\\
		216& 9741& F814W& 2800\\
		382& 8163& F814W& 85\\
		385& 8163& F814W& 85\\
		392& 8163& F814W& 85\\
		398& 8163& F814W& 85\\
		470& 8163& F814W& 85\\
		488& 8163& F814W& 85\\
		537& 8163& F814W& 85\\
		577& 8163& F814W& 85\\
		581& 8163& F814W& 85\\
		682& 8163& F673N& 2000\\
		706& 5114& F814W& 1200\\
		849& APPP-SFD& F606W& 1760 \\
		\hline 
	\end{tabular}
	\begin{flushleft}
		{{\bf Notes.} The completeness limit tests that we conducted were done on VFTS 385, which is amongst the shortest exposure times. This means that our completeness limit of 2 to 3 \msun\ is a worst-case scenario. The HST ID column refers to the proposal IDs when available, otherwise they refer to the APPP repository \citep{Wadadekar2006}.} 
	\end{flushleft}
\end{table}

\section{Probability test for chance alignment of runaway stars with filaments} \label{sec:appendix} Currently, there are no high-resolution, continuous, large field-of-view maps of the 30 Doradus region that show where the star forming filamentary gas and dust are. In the coming year this deficiency will be resolved as the\emph{ Herschel Space Observatory} survey maps of the LMC, HERschel Inventory of The Agents of Galaxy Evolution (HERITAGE) \citep{Meixner2010}, will be publicly released. By combining the HERITAGE maps with the data presented in this paper one could reject the likelihood of O-stars being runaways based on probability. We propose that the definition of a filament from Herschel observations is best defined by the 2nd differential maps \cite{Molinari2010} introduced, where a physically relevant radius from the peaks of the filaments can bear association to the stars.

\begin{figure}
	\includegraphics[width=8.5cm]{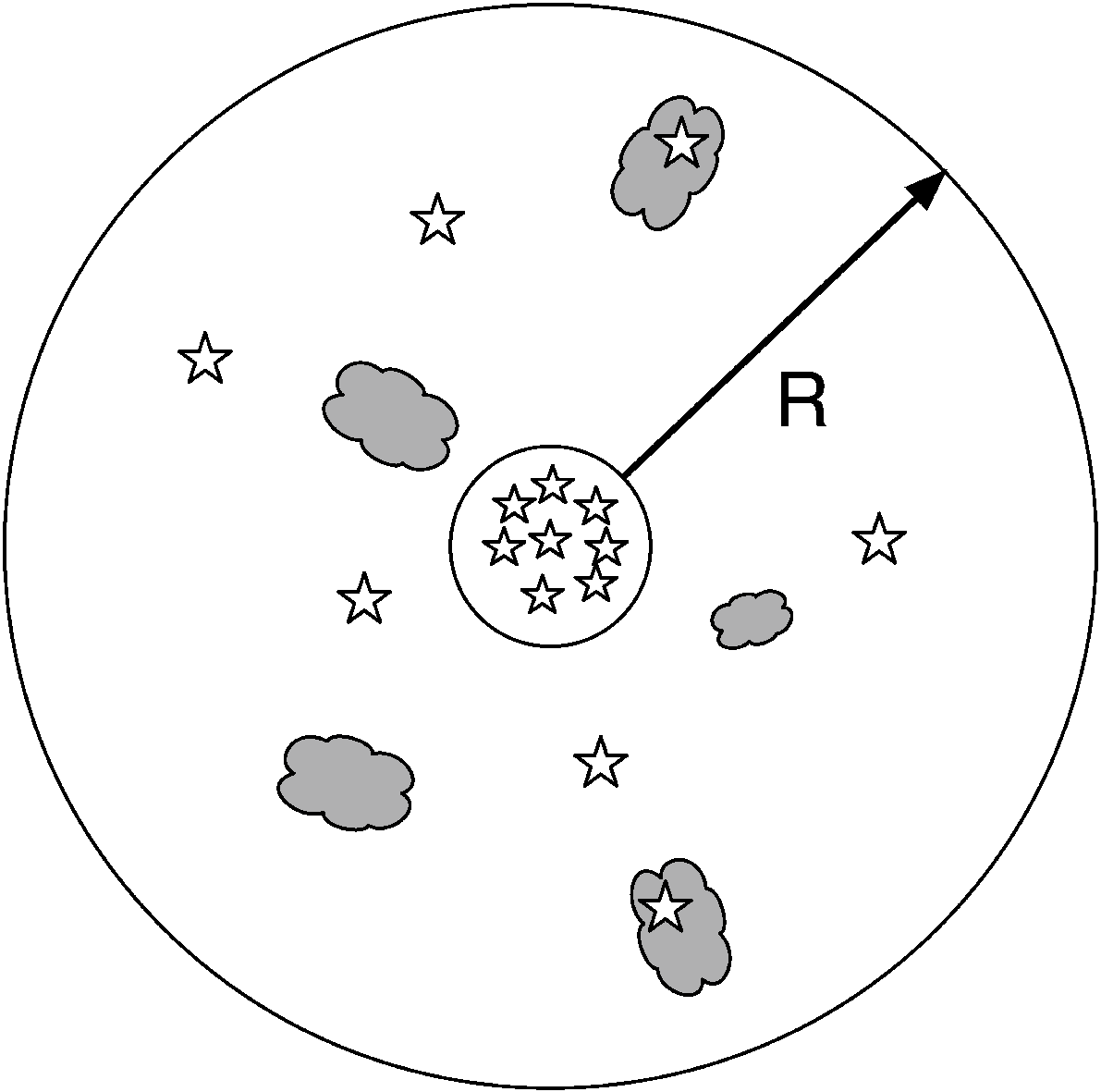} \caption{A diagram showing how the probabilistic method should be conceptualised. The radius $R = R_{\textrm{boundary}}-R_{\textrm{cluster}}$. All the stars outside of the cluster are initially assumed to be runaways for the binomial probability problem to calculate the likelihood of multiple alignment events between the runaways and the filaments (grey clouds). The filaments at these scales, $> 5~pc$, will not be affected by a single O-star such that nothing remains.} \label{fig:figb01} 
\end{figure}

To do this, we take O-stars similar to the ones discussed in this paper and assume that they are runaways. Using the radial velocity measurements from the Tarantula Survey, we approximate that there are roughly $N$ number of runaways in total. Then we place a circular boundary around R136 at distance $R$, where $R = R_{\textrm{boundary}}-R_{\textrm{cluster}}$, and has area $A_{\textrm{boundary}}$. Within $R$ distance, let's assume that gas/dust filaments covers an area $A_{\textrm{filament}}$, such that $A_{\textrm{filament}} \le A_{\textrm{boundary}}$. Refer to Fig.~\ref{fig:figb01} for a visual context of the scenario we have now discussed.

With the above conditions, what is the probability that $n$ of $N$ runaway O-stars line up with filamentary structures in the line-of-sight? For simplicity, assume that the $N$ O-stars are randomly distributed within the boundary such that the probability of an O-star lining up with a filament can be expressed as $\rho$.
\begin{equation}
	\rho = \frac{A_{\textrm{filament}}}{A_{\textrm{boundary}}} 
\end{equation}

To see what probability $n$ of $N$ ($n \le N$) O-stars is in the line-of-sight with filaments we invoke the binomial probability equation as shown below, where we assume that each line-of-sight event is independent of one another.
\begin{equation}
	P(n) = {N \choose n}\rho^n(1-\rho)^{N-n} 
\end{equation}

\bibliographystyle{aa} 
\bibliography{jabref}

\end{document}